\documentclass[a4paper,11pt]{article}
\pdfoutput=1 
\usepackage{amssymb,amsmath,amsfonts,makeidx,placeins,tikz}
\usepackage{graphicx,rotate,subcaption,color,slashed,caption,epstopdf}
\usepackage[colorlinks=true,
            linkcolor=magenta,
            urlcolor=blue,
            citecolor=blue]{hyperref}
\numberwithin{equation}{section}
\numberwithin{figure}{section}

\evensidemargin=\oddsidemargin
\begin{document}

\begin{center}

{\Large \bf Improving Fine-tuning in Composite Higgs Models}\\
\vspace*{0.5cm} {\sf Avik
  Banerjee~$^{a,}$\footnote{avik.banerjeesinp@saha.ac.in}, ~Gautam
  Bhattacharyya~$^{a,}$\footnote{gautam.bhattacharyya@saha.ac.in}, ~Tirtha
  Sankar Ray~$^{b,}$\footnote{tirthasankar.ray@gmail.com}} \\
\vspace{10pt} {\small } $^{a)}$ {\em Saha Institute of Nuclear
    Physics, HBNI, 1/AF Bidhan Nagar, Kolkata 700064, India}
\\
\vspace{3pt} {\small } $^{b)}${\em Department of Physics and Centre
  for Theoretical Studies, Indian Institute of Technology Kharagpur,
  Kharagpur 721302, India} \normalsize
\end{center}

\begin{abstract}
In this paper we investigate the next-to-minimal composite Higgs model
with a $\rm{SO(6)}/\rm{SO(5)}$ coset, whose pNGB sector includes a
Standard Model singlet in addition to the usual Higgs doublet.  The
fermions are embedded in the representation $\bf 6$ of
$\rm{SO(6)}$. We study the region of parameter space of the model
where the radiatively generated potential has global minima with both
the doublet and the singlet fields developing vacuum expectation
values.  We investigate the consequences of kinetic and mass mixing
between the Higgs and the singlet scalar that arise in this
framework. We demonstrate that the ensuing doublet-singlet mixing can
provide a handle to accommodate heavier resonances (top-partners) for
a given compositeness scale as compared to the minimal composite Higgs
model, thus relaxing the tension with the direct LHC bounds. The main
phenomenological consequence of this is a sizable deviation of the
Higgs couplings from the Standard Model predictions. While the present
experimental precision in the measurement of the Higgs couplings still
allows for considerable release of this tension, future measurements
of the Higgs branching ratios with increased precision would lead to
stringent constraints on this setup.
\end{abstract}


\bigskip

\section{Introduction}
\label{intro}
The continued absence of new physics at experiments in the energy and
intensity frontiers complemented with lack of any pointers from
astrophysical experiments have squeezed the space for TeV scale beyond
Standard Model physics (BSM) to claustrophobic proportions. However,
several theoretical issues like the stability of the weak scale from
any high scale dynamics that the Standard Model (SM) might couple to,
or the existence of the hyper-charge Landau pole
\cite{Giudice:2014tma} at a trans-Planckian scale demanding nontrivial
dynamics at higher scales, indicate that the SM is an effective
theory. These and associated theoretical issues remain the main
driving force behind the BSM physics. As we enter the precision Higgs
era following discovery of the Higgs boson
\cite{Aad:2012tfa,Chatrchyan:2012xdj} at $\sim 125$ GeV, in addition
to the electroweak precision constraints a set of new constraints from
Higgs physics gets imposed on BSM scenarios.

The composite Higgs framework, where the Higgs is identified with a
pseudo-Nambu-Goldstone boson (pNGB) originating from spontaneous
breaking of a global symmetry in a strongly interacting sector,
provides a consistent framework to shield the weak scale from the
gauge hierarchy problem
\cite{Kaplan:1983fs,Dugan:1984hq,Contino:2010rs,Panico:2015jxa,Csaki:2016kln,
  Giudice:2007fh}.  The scale of spontaneous symmetry breaking
identified as the compositeness scale separates the weak scale from
any higher dynamics.  It is the related inexact shift symmetry of the
pNGB Higgs  that protects the Higgs
mass from sensitivity to the UV scale. In these models the Higgs
potential is generated at one-loop by the explicit breaking of the
global symmetries of the strong sector that is communicated by the
linear mixing of the strong sector operators with the SM
states. Within this framework of the so called \textit{partial
  compositeness,} where the potential is
generated mainly by top quark induced interactions, the Higgs mass is
expected to scale as \cite{Pomarol:2012qf}
\begin{equation}
\label{nmchm_s1:1}
m_h^2 \sim \frac{N_c}{\pi^2}\frac{m_t^2 m_Q^2}{f^2} \sim
\frac{N_c}{\pi^2}y_t^2 \frac{m_Q^2}{\Delta}~,
\end{equation}
where $m_Q$ is a generic mass of the strong sector resonances that mix
with the top and $f$ is the compositeness scale. The ratio $\Delta
\equiv\xi^{-1}\equiv f^2/v^2_{ew}$ is a measure of tuning required to
obtain the electroweak vacuum expectation value (vev) of the Higgs
$(\mathnormal{v}_{ew})$ as compared to the compositeness scale $f$. In
general, the vev-tuning in this class of models is expected to be
greater than $\Delta$.  In most cases it can be estimated as $\sim
\Delta / \kappa$, where $\kappa (\lesssim 1)$ is a model dependent parameter
\cite{Panico:2012uw, Archer:2014qga, Barnard:2015ryq, Csaki:2017cep,
  Barnard:2017kbb}. In this paper we assume that $\Delta$ quantifies
the {\em minimal} tuning in the Higgs vev. In fact, we mainly focus on
relative tuning between models, where the numerical impact of $\kappa$
mostly cancels out in ratios.  It is clear from Eq.~\eqref{nmchm_s1:1}
that the relative lightness of the Higgs boson requires either a
relatively light coloured resonance or a large fine-tuning. The
non-observation of any exotic resonance at the LHC
\cite{ATLAS-CONF-2016-032} implies larger values of $\Delta$ and hence
more fine-tuned scenario. This is basically a restatement of the more
generic observation that the measured Higgs boson mass of $\sim 125$
GeV is somewhat on the lower side for the otherwise well-motivated
composite Higgs framework.  This may be contrasted with the
supersymmetric extension of the SM where the Higgs mass is perceived
to be on the heavier side \cite{Gherghetta:2012gb, Brummer:2012ns}.

The connection of light resonances with composite Higgs mass has been
studied extensively in the context of minimal composite Higgs model
(MCHM) based on $\rm{SO(5)}/\rm{SO(4)}$ coset applying the QCD-like
Weinberg sum rules
\cite{Contino:2006qr,Azatov:2011qy,Pomarol:2012qf,Panico:2012uw,Matsedonskyi:2012ym},
effective two-site models \cite{Panico:2012uw} or from explicit
calculations with the 5D duals of these theories \cite{Agashe:2004rs,
  Agashe:2005dk}. Within the MCHM framework the LHC results
\cite{ATLAS-CONF-2016-032} put severe constraints on the region of
parameter space with moderate
$\Delta\sim\mathcal{O}(10)$. Interestingly, it was shown that two-loop
contributions to the Coleman-Weinberg (C-W) potential from the
coloured vector resonances of the strong sector relaxes this by $5-10
\%$ \cite{Barnard:2013hka}.  Implications of the lepton ($\tau$)
resonances on fine-tuning have also been considered
\cite{Carmona:2014iwa,Carmona:2013lva}.

\begin{figure}[t]
\centering
\includegraphics[trim = 20mm 240mm 20mm 15mm, clip,width=\textwidth]{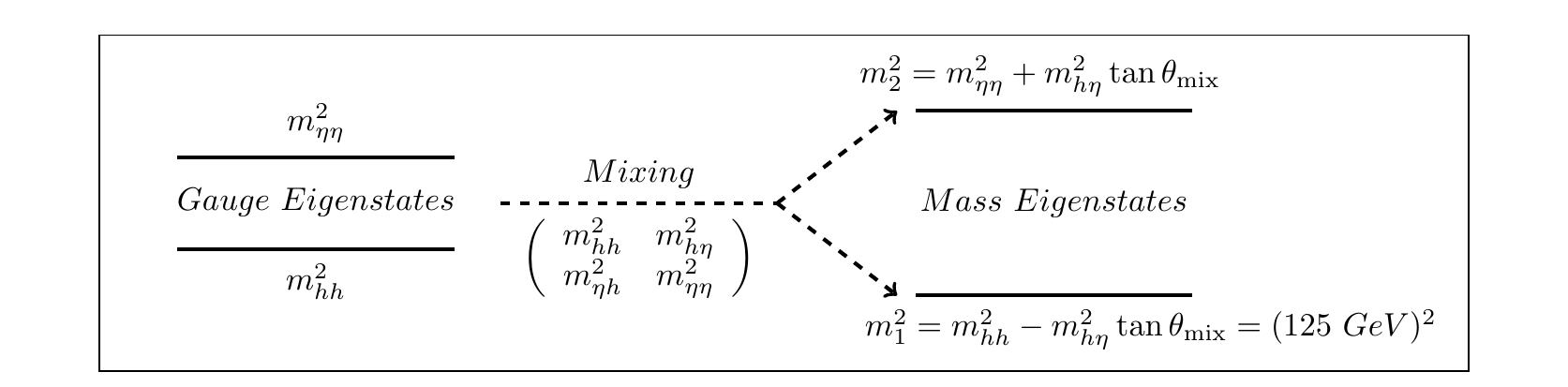}
\caption{\small\it Schematic diagram describing level-repulsion.}
\label{nmchm_s3:fig}
\end{figure}

In this paper we explore the possibility of increasing the mass gap
between the top-partner resonances and the Higgs boson for a given
$\Delta$ by employing a possible tree-level doublet-singlet mixing in
the pNGB scalar sector of non-minimal model.  If the singlet state
is heavier, the mixing can lead to a {\em level-repulsion} pushing the
dominantly doublet eigenstate down to match the observed Higgs mass at
125 GeV. The masses of both the states before mixing are conceivably
larger and hence more natural from the composite Higgs perspective.
The setup is depicted schematically in Fig.~\ref{nmchm_s3:fig}. This
possibility naturally demands a larger set of pNGBs in the strong
sector than in MCHM, implying an enlargement of the coset space. Many
such non-minimal composite Higgs frameworks have been discussed in the
literature \cite{Gripaios:2009pe, Galloway:2010bp, Mrazek:2011iu,
  Chala:2012af, Bertuzzo:2012ya, Chala:2016ykx}.

Here we will confine ourselves to the next-to-minimal composite Higgs
model with a coset $\rm{SO(6)}/\rm{SO(5)}$, which would suffice to
demonstrate the phenomenon of level-repulsion as a proof of principle.
This coset represents the minimal extension beyond MCHM, as it
introduces an extra SM gauge-singlet scalar along with the four
components of the usual Higgs doublet
\cite{Gripaios:2009pe,Redi:2012ha,Serra:2015xfa,Low:2015qep,Cai:2015bss,
  Arbey:2015exa}. Several aspects of this model has been discussed in
the literature, e.g.  in the context of dark matter
\cite{Frigerio:2012uc,Marzocca:2014msa,Fonseca:2015gva,Kim:2016jbz}
and electroweak baryogenesis \cite{Espinosa:2011eu}. Incidentally this
also represents the minimal coset that allows for a 4D ultraviolet
completion
\cite{Katz:2005au,Barnard:2013zea,Ferretti:2013kya,Cacciapaglia:2014uja,
  Agugliaro:2016clv,Galloway:2016fuo}. Here we will be interested in
the parameter space where the radiatively generated C-W potential for
the Higgs doublet and the SM singlet minimizes to non-zero vevs for
both scalars. This may allow for a nontrivial mixing between the Higgs
and the SM singlet in both the kinetic terms and the potential.  This
is somewhat complementary to most of the studies carried out in the
literature for the non-minimal models where the singlet is assumed not
to develop a vev making it a possible dark matter candidate. The
collider phenomenology of this framework has recently been studied
\cite{Niehoff:2016zso}.

The paper is organised as follows. In section \ref{s2}, we briefly
review the MCHM, specifically focusing on the r\^ole of top-partner
resonances for reproducing the 125 GeV mass of the Higgs boson. In
section \ref{s3}, we review the next-to-minimal model with
$\rm{SO(6)}/\rm{SO(5)}$ coset and the associated C-W potential.  In
section \ref{imp-fine}, we discuss the relaxation in the correlation
between the top-partner and the Higgs mass induced by the level-splitting mechanism. Some phenomenological implications are discussed
in section \ref{pheno}, before concluding in section \ref{s4}.

\section{Higgs mass and light top-partners in minimal model}
\label{s2}
In this section we briefly review the C-W scalar potential and the
correlation between the light top-partners and the Higgs mass within
MCHM.  Its coset is $\rm{SO(5)}/\rm{SO(4)}$ having four pNGBs that can
be identified with the components of the SM Higgs doublet. The pNGBs
can be parametrised with the non-linear representation ($\Sigma$)
\cite{Callan:1969sn,Coleman:1969sm} in the unitary gauge in terms of
the dimensionless field $h$ normalised by the compositeness scale
($f$) as \cite{Agashe:2004rs}
\begin{equation}
\label{nmchm_s2:2}
\Sigma = \left( 0,~0,~0,~h,~\sqrt{1-h^2} \right)^T.
\end{equation}

The Higgs potential and the phenomenology of the model depends on the
representation into which the SM fermions are embedded.  The minimal
representation that ensures a custodial protection of the $Z
b_L\bar{b}_L$ coupling is the fundamental $\bf 5$ of SO(5). Here we
confine our discussion to this minimal framework $\rm{MCHM}_5$. The
most important contribution to the Higgs potential driving electroweak
symmetry breaking (EWSB) comes from the top sector. The relevant part
of the Lagrangian is given by \cite{Pomarol:2012qf}
\begin{eqnarray}
\label{nmchm_s2:3}
\mathcal{L}=\overline{t}_L\slashed{p}\left[
  \Pi^{t_L}_0+\frac{\tilde{\Pi}^{t_L}_1}{2}h^2\right]t_L+\overline{t}_R\slashed{p}\left[
  \Pi^{t_R}_0+\tilde{\Pi}^{t_R}_1(1-h^2)\right]t_R+\overline{t}_L
\left[\frac{M^{t}}{\sqrt{2}}h\sqrt{1-h^2}\right]t_R
+\rm{h.c.} \, ,
\end{eqnarray}
where the $\Pi$'s are the structure functions which depend on the
top-partner masses of the strong sector.  The one-loop C-W potential
has the structure \cite{Panico:2015jxa,Marzocca:2012zn}
\begin{equation}
\label{nmchm_s2:4}
V_{\rm{eff}}(h)= -\frac{\mu^2}{2} h^2+\frac{\lambda}{4}h^4~. 
\end{equation} 
Minimization of the potential in Eq.~\eqref{nmchm_s2:4} yields
\begin{equation}
\label{nmchm_s2:5}
\xi \equiv \langle h\rangle^2 =\frac{\mu^2}{\lambda} ~.
\end{equation} 
The Higgs mass can be expressed in terms of the coefficient $\lambda$
and the parameter $\xi$ as
\begin{equation}
\label{nmchm_s2:6}
m_h^2=\frac{2}{f^2}\xi(1-\xi)\lambda ~.
\end{equation} 
To calculate the parameters of the potential we utilise the Weinberg
sum rules to model the structure functions. We employ two strong
sector resonances (one singlet and a quadruplet under $\rm SO(4)$)
with masses $m_{Q_1}$ and $m_{Q_4}$ which saturate the
integrals. Calculation of the coefficients $\mu^2$ and $\lambda$ from
the top and gauge sectors is described in Appendix \ref{appendix
  minimal model}.

 In Fig.~\ref{nmchm_s2:f1} we plot the contours for $m_h = 125$ GeV in
 the $m_{Q_1}-m_{Q_4}$ plane for $\xi =0.08$ and $0.04$. The figure
 suggests that a light Higgs boson implies at least one of the
 top-partners is relatively light. The present LHC constraints on the
 top-partners exclude the region below $(m_{Q_1},~m_{Q_4})\sim 1$ TeV
 \cite{ATLAS-CONF-2016-032}. The situation changes when one embeds the
 fermions in other representations, see for example
 \cite{Pappadopulo:2013vca}. Here we make a passing remark on the
 issue of double-tuning \cite{Matsedonskyi:2012ym, Panico:2012uw},
 which eases in cases when there are multiple invariants in the Yukawa
 structure. In the presence of double-tuning, the total fine-tuning in
 the Higgs vev can be estimated by $\Delta/\kappa$, as mentioned
 previously\footnote{A complementary method based on statistical
   approach to estimate the fine-tuning can be found in
   \cite{Barnard:2017kbb}.}. In case of $\rm MCHM_5$, $\kappa$ can be
 na\"ively parametrised as $\kappa \sim (|F_Q|/m_{Q})^2$, where $F_Q$
 and $m_Q$ represent the decay constant and mass of the lightest
 fermionic resonances, respectively. For illustration, a typical
 estimate shows $\kappa \sim 0.3$ with the resonance mass $m_Q \le
 1.5$ TeV and $\Delta = 10$ \cite{Panico:2012uw}.

\begin{figure}[t]
\centering
\includegraphics[width=100mm]{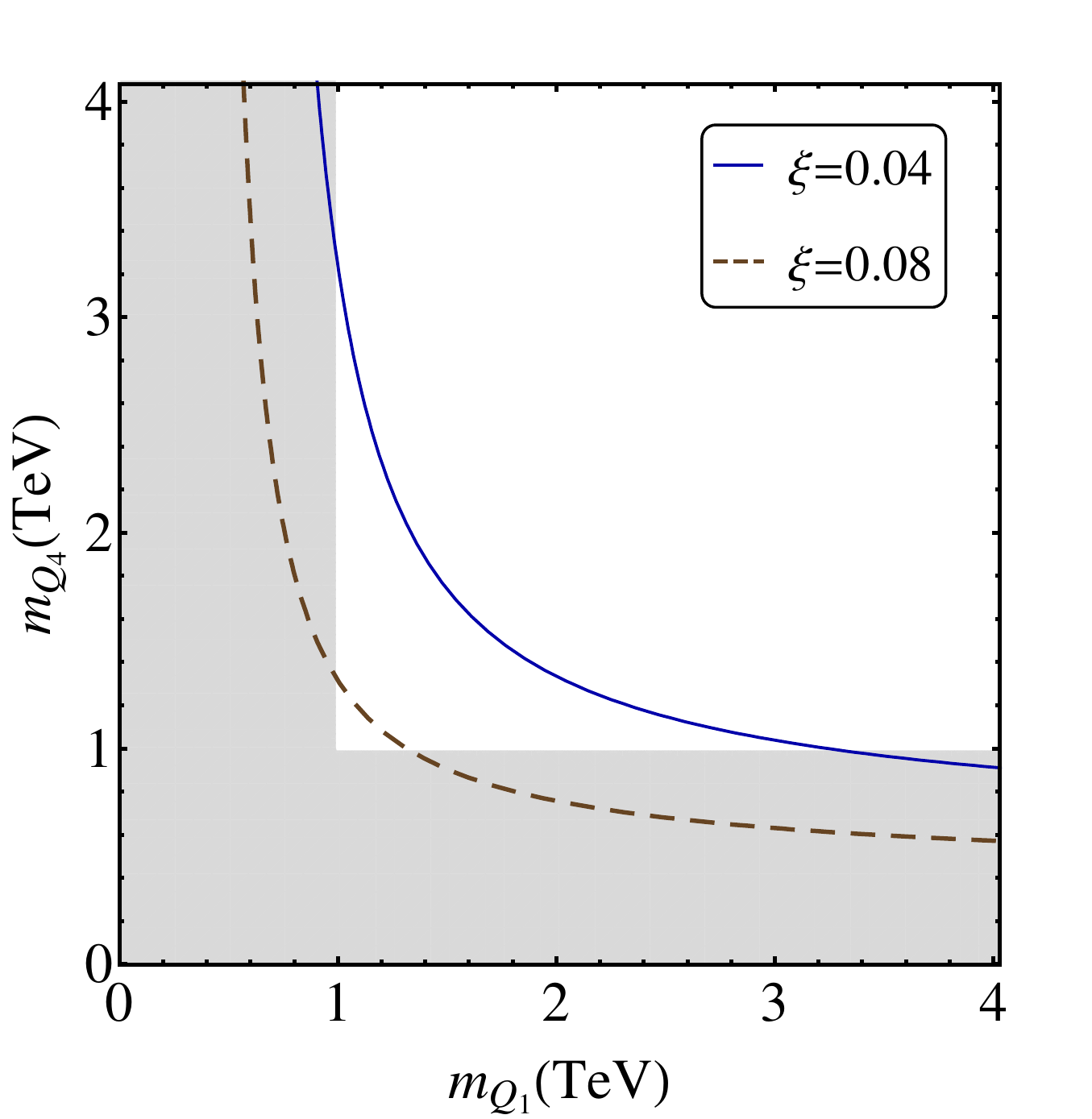}
\caption{\small\em The 125 GeV Higgs mass contour in $m_{Q_1}-m_{Q_4}$
  plane is displayed for two different choices of $\xi$. The blue curve
  corresponds to $\xi=0.04$ (i.e. $\Delta=25$), and the brown dashed
  line corresponds to $\xi=0.08$ (i.e. $\Delta\simeq 13$). The gray
  area is already excluded from LHC direct searches.}
\label{nmchm_s2:f1}
\end{figure}

\section{Next-to-minimal  model: $\rm{SO(6)}/\rm{SO(5)}$ coset}
\label{s3}

The next-to-minimal model enlarges the coset to
$\rm{SO(6)}/\rm{SO(5)}$ \cite{Gripaios:2009pe, Cacciapaglia:2014uja,
  Franzosi:2016aoo, Serra:2015xfa} which includes an additional CP-odd
SM singlet along with the usual pNGB Higgs doublet. Interestingly this
coset is homomorphic to $\rm{SU(4)}/\rm{Sp(4)}$.  A brief review of
the salient features of this model is now in order.

\subsection{Parametrization of pNGB degrees of freedom}
\label{s3.1}
The spontaneous breaking of global $\rm{SO(6)} \to \rm{SO(5)}$, at the
composite scale $f$, delivers five pNGBs.
Four of them transform as $\bf(2,2)$ under $\rm{SO(4)}\simeq
\rm{SU(2)}_L\times \rm{SU(2)}_R$, and the other is a singlet of
$\rm{SO(4)}$ transforming as $\bf(1,1)$. The pNGBs are parametrised as
\begin{equation}
\label{nmchm_s3:1}
\Sigma
=e^{i\frac{\sqrt{2}}{f}\pi^{\hat{\alpha}}\hat{T}^{\hat{\alpha}}}
\Sigma_0=\frac{1}{\pi}\sin\frac{\pi}{f}\left(\pi_1,~\pi_2,~\pi_3,~\pi_4,~\pi_5,~
\pi \cot\frac{\pi}{f} \right)^T ,
\end{equation}
where $\hat{T}^{\hat{\alpha}}$ are the generators along the broken
directions as given in Appendix \ref{appendix A}, and $\Sigma_0$
denotes the vacuum $(0,0,0,0,0,1)^T$. In unitary gauge, $\Sigma$ can
be written in terms of a CP-even field, $h \equiv
(\pi_4/\pi)\sin(\pi/f)$, and a CP-odd field, $\eta \equiv
(\pi_5/\pi)\sin(\pi/f)$, as \cite{Frigerio:2012uc}
\begin{equation}
\label{nmchm_s3:4}
\Sigma=\left( 0,~0,~0,~h,~\eta,~\sqrt{1-h^2-\eta^2} \right)^T.
\end{equation}

\subsection{Gauge sector and kinetic mixing}
\label{s3.2}
The kinetic term for the  pNGBs can be written as
\begin{equation}
\label{nmchm_s3:5}
\mathcal{L}_{\rm{kinetic}}=\frac{f^2}{2}(D_\mu\Sigma)^T(D_\mu\Sigma) ~,
\end{equation}
where the covariant derivative is
$D_\mu=(\partial_\mu-igT^a_LW^a_\mu-ig'T^3_RB_\mu)$, with $g$ and $g'$
as the SM gauge couplings. Expanding the covariant derivatives, we
get the complete gauge-kinetic term as
\begin{eqnarray}
\label{nmchm_s3:6}
\mathcal{L}_{\rm{kinetic}}=\frac{f^2}{2}\bigg[(\partial_\mu h)^2+
  (\partial_\mu \eta)^2+\frac{(h\partial_\mu
    h+\eta\partial_\mu\eta)^2}{1-h^2-\eta^2}+\frac{g^2h^2}{2}\bigg(|W|^2+\frac{1}{2\cos^2\theta_w}Z^2\bigg)\bigg]
~.~
\end{eqnarray}
In the region of parameter space where both $h$ and $\eta$ get vevs, a
kinetic mixing between $h$ and $\eta$ is obtained. This becomes
explicit once we write down the Lagrangian in terms of the shifted
fields (i.e. $h\rightarrow h+\langle h\rangle$ and $\eta\rightarrow
\eta+\langle\eta\rangle$)
\begin{eqnarray}
\nonumber
\mathcal{L}_{\rm{kinetic}} \supset \frac{f^2}{2}\left[\left(1+\frac{\langle
    h \rangle^2}{1-\langle h
    \rangle^2-\langle\eta\rangle^2}\right)(\partial_\mu h)^2+
  \left(1+\frac{\langle \eta \rangle^2}{1-\langle h
    \rangle^2-\langle\eta\rangle^2}\right)(\partial_\mu
  \eta)^2\right.\\
\label{nmchm_s3:7}
+\left.\left(\frac{2\langle h \rangle\langle\eta\rangle}{1-\langle h
  \rangle^2-\langle\eta\rangle^2}\right)(\partial_\mu h)(\partial_\mu
\eta)\right] ~.~~
\end{eqnarray} 
The kinetic term can be canonically normalised by a non-unitary
rotation of $h$ and $\eta$, as is routinely employed in Radion-Higgs
scenarios \cite{Chaichian:2001rq,Dominici:2002jv,Desai:2013pga} or in
the case of kinetic mixing in abelian gauge extensions of the SM
\cite{Babu:1997st}. The canonically normalised fields denoted by
$\{h_n,\eta_n\}$ are related to the gauge eigenstates by the following
redefinitions
\begin{equation}
\label{nmchm_s3:8}
\begin{split}
h=a h_n,
\qquad
\eta=b h_n+c\eta_n ~,
\end{split}
\end{equation}
where the coefficients  are given by
\begin{equation}
\label{nmchm_s3:9}
\begin{split}
a=\frac{1}{f}\sqrt{1-\langle h \rangle^2}, \qquad
b=-\frac{1}{f}\frac{\langle h\rangle
  \langle\eta\rangle}{\sqrt{1-\langle h\rangle^2}}, \qquad
c=\frac{1}{f}\frac{\sqrt{1-\langle h
    \rangle^2-\langle\eta\rangle^2}}{\sqrt{1-\langle h \rangle^2}} ~.
\end{split}
\end{equation}
The transformation described in Eq.~\eqref{nmchm_s3:8} is not
unique\footnote{The general non-unitary rotation is of the form,
\begin{equation}
\nonumber
\begin{split}
h=a h_n+d\eta_n~,
\qquad
\eta=b h_n+c\eta_n ~.
\end{split}
\end{equation}
To calculate the coefficients, we need to put the above
transformations back in Eq.~\eqref{nmchm_s3:7} and demand that,
coefficients of $(\partial_\mu h_n)^2$ and $(\partial_\mu \eta_n)^2$
would be $1/2$ and that of $(\partial_\mu h_n)(\partial_\mu \eta_n)$
would be zero. Definitely, there are three constraint equations and
four variables $a,b,c,d$. Therefore, no unique solution can be
found. We take the special choice $d=0$ for our analysis. We
have checked that the final results do not depend on the choice of the
shift.}, however the phenomenology of the ensuing theory is
independent of this choice.

\subsection{Fermion embedding}
\label{s3.3}

The radiatively generated potential receives contributions from the
gauge and Yukawa sectors. Gauge interactions generate a potential only
for the doublet state $h$. Further, the potential for $\eta$ is
protected by a global $\rm{U(1)}_\eta$ symmetry, which is isomorphic
to a $\rm{SO(2)}$ rotation in the 5-6 direction of
Eq.~(\ref{nmchm_s3:4}) \cite{Gripaios:2009pe,Frigerio:2012uc}. To
generate a potential for $\eta$, this $\rm{U(1)}_\eta$ has to be
broken explicitly by the SM Yukawa couplings. This implies that some
SM fermions embedded in an incomplete multiplet of $\rm{SO(6)}$ should
be charged under $\rm{U(1)}_\eta$.  In the present analysis, we embed
the SM fermions in the fundamental {\bf 6} representation of SO(6).
Under the group decomposition $\rm{SO(6)}\supset \rm{SO(4)}\times
\rm{SO(2)}\simeq \rm{SU(2)}_L\times \rm{SU(2)}_R\times
\rm{U(1)}_\eta$, one obtains
\begin{eqnarray}
\label{nmchm_s3:12}
\mathbf{6}=(\mathbf{2},\mathbf{2})_0\oplus
(\mathbf{1},\mathbf{1})_2\oplus (\mathbf{1},\mathbf{1})_{-2} ~,
\end{eqnarray}
where the subscripts denote the $\eta$-charge. Note that the two
singlets under $\rm{SO(4)}\simeq \rm{SU(2)}_L\times \rm{SU(2)}_R$ in
Eq.~\eqref{nmchm_s3:12} are charged under $U(1)_\eta$ and hence are
capable of breaking the symmetry protecting $\eta$. The left-handed
top quark is embedded into $\bf(2,2)$ protecting the $Zb_L\bar{b_L}$
coupling \cite{Agashe:2006at}. On the other hand, right-handed top
quark is embedded as a linear combination in both $\bf(1,1)$. To
generate proper SM hypercharges, we assume that the representation
{\bf 6} of SO(6) is additionally charged under a global ${\rm U(1)_X}$ with
$X=2/3$, and $Y = T^3_R + X$. Finally, we write down
the explicit embeddings for the top quark as
\begin{eqnarray}
\label{nmchm_s3:13}
Q_L&=&\frac{1}{\sqrt{2}}(-ib_L,~-b_L,~-it_L,~t_L,~0,~0)^T ,\\
\label{nmchm_s3:14}
T_R&=&(0,~0,~0,~0,~e^{i\delta}c_\theta t_R,~s_\theta t_R)^T ,
\end{eqnarray}
where $c_\theta (s_\theta)$ denote the cosine (sine) of an angle
$\theta$ which is a free parameter \cite{Redi:2012ha}.  Clearly,
$\theta=\pi/4$ and $\delta=\pi/2$, restores the $\rm{U(1)}_\eta$ in
the sense that potential for $\eta$ is no longer generated from the
top sector making it an electroweak axion which is severely
constrained from charged kaon decay \cite{Georgi:1986df}.

\subsection{Effective potential}
With the top quark being embedded in the fundamental representation of
$\rm{SO(6)}$ as described in Eqs.~\eqref{nmchm_s3:13} and
\eqref{nmchm_s3:14} and the nonlinear realization of pNGBs given in
Eq.~\eqref{nmchm_s3:4}, one can write down the effective Lagrangian
for the top-Higgs sector in terms of the group theoretic
invariants. We assume $\delta = \pi/2$ in Eq.~\eqref{nmchm_s3:14},
which considerably simplifies the derived potential without losing any
key feature required for the present discussion. The effective
Lagrangian is given by
\begin{eqnarray}
\nonumber
\mathcal{L}=\overline{t}_L\slashed{p}\left[ \Pi^{t_L}_0+\frac{\tilde{\Pi}^{t_L}_1}{2}h^2\right]t_L+\overline{t}_R\slashed{p}\left[ \Pi^{t_R}_0+\tilde{\Pi}^{t_R}_1\left( c_{2\theta}\eta^2+s_\theta^2(1-h^2)\right)\right]t_R\\
\label{nmchm_s3:15}
+\overline{t}_L\left[\frac{M^{t}}{\sqrt{2}}h\left(
  ic_\theta\eta+s_\theta\sqrt{1-h^2-\eta^2}\right)\right]t_R
+\rm{h.c.}
\end{eqnarray}
As anticipated, the couplings to $\eta$ are through $t_R$ alone as it
has a non trivial $\eta$-charge.  The effective one-loop
potential involving the dimensionless fields $h$ and $\eta$ obtained
from the above Lagrangian can be parametrised as
\begin{equation}
\label{nmchm_s3:16}
V_{\rm{eff}}(h,\eta)=-\frac{\mu_1^2}{2}h^2+\frac{\lambda_1}{4}h^4-\frac{\mu_2^2}{2}\eta^2+\frac{\lambda_2}{4}\eta^4-\frac{\lambda_m}{2}h^2\eta^2 ~.
\end{equation}
A detailed calculation of the potential from partial compositeness
scenario for the coefficients $\mu_1,~\lambda_1,~\mu_2,~\lambda_2$ and
$\lambda_m$ are given in Appendix \ref{appendix B} using the Weinberg
sum rules framework.  The convergence of integrals involved in $\mu_1$
and $\mu_2$ requires introduction of at least three top-partner
resonances, whereas for the calculability of the other three
coefficients only two resonances would suffice. The coefficients
$\mu_1$ and $\lambda_1$ contain gauge as well as top contributions,
whereas the rest of the parameters contain only the top contributions.
This implies that a generic minimization is expected to yield a
non-zero vev for $\eta$ arising from the negative contribution of the
top sector to the $\eta$ quadratic.  On the other hand, the vev of the doublet
requires a smart cancellation between the top and gauge sectors. The
minimum of the above potential corresponds to
\begin{equation}
\label{nmchm_s3:17}
\begin{split}
\xi \equiv \langle
h\rangle^2=\frac{\lambda_2\mu_1^2+\lambda_m\mu_2^2}{\lambda_1\lambda_2-\lambda_m^2}~,
\qquad \chi \equiv \langle
\eta\rangle^2=\frac{\lambda_1\mu_2^2+\lambda_m\mu_1^2}{\lambda_1\lambda_2-\lambda_m^2}
~.
\end{split}
\end{equation}
Recall, $\xi \equiv v_{\rm ew}^2/f^2$, where $v_{\rm ew} = 246$ GeV.
Note, $\lambda_1,~\lambda_2>0$ and $\lambda_1\lambda_2-\lambda_m^2>0$
ensure stability of the potential.  The condition for both $h$ and
$\eta$ to develop vevs implies $\mu_1^2,\mu_2^2 >0$. Also it follows
from Eq.~\eqref{nmchm_s3:4} that
\begin{equation}
\xi+\chi\le 1 ~.
\end{equation}
In terms of the canonically normalised fields we obtain the
following scalar mass matrix
\begin{equation}
\label{nmchm_s3:20}
M^2(h_n,\eta_n)=\left(\begin{array}{cc}
m_{h_nh_n}^2 & m_{h_n\eta_n}^2 \\ 
m_{\eta_n h_n}^2 & m_{\eta_n\eta_n}^2
\end{array}\right) ~;
\end{equation}
where
\begin{eqnarray}
\label{nmchm_s3:21}
m_{h_nh_n}^2 &=& 2\lambda_1a^2\xi+2\lambda_2b^2\chi-4\lambda_mab\sqrt{\xi\chi} ~,\\
\label{nmchm_s3:22}
m_{\eta_n\eta_n}^2 &=& 2\lambda_2c^2\chi ~,\\
\label{nmchm_s3:23}
m_{h_n\eta_n}^2&=&m_{\eta_nh_n}^2 = 2\lambda_2bc\chi-2\lambda_mac\sqrt{\xi\chi} ~.
\end{eqnarray}
To avoid tachyonic eigenvalues we impose $m_{h_nh_n}^2,
m_{\eta_n\eta_n}^2 >0$ and ${Det}[M^2]> 0$. Clearly a non-zero
$m_{h_n\eta_n}^2$ gives rise to mass-mixing between the doublet and
singlet. The mass eigenvalues can be calculated as \cite{Jeong:2012ma}
\begin{eqnarray}
\label{nmchm_s3:24}
m_{\hat{\eta}}&=&\sqrt{m_{\eta_n\eta_n}^2+m_{h_n\eta_n}^2\tan\theta_{\rm{mix}}} ~,\\
\label{nmchm_s3:25}
m_{\hat{h}}&=&\sqrt{m_{h_nh_n}^2-m_{h_n\eta_n}^2\tan\theta_{\rm{mix}}} ~,
\end{eqnarray}
where the doublet-singlet mixing angle $\theta_{\rm{mix}}$ is given by
\begin{equation}
\label{nmchm_s3:26}
\tan2\theta_{\rm{mix}}=\frac{2m_{h_n\eta_n}^2}{m_{\eta_n\eta_n}^2-m_{h_nh_n}^2} ~.
\end{equation}
The eigenvectors corresponding to the eigenvalues $m_{\hat{\eta}}$ and
$m_{\hat{h}}$ are given by
\begin{eqnarray}
\label{nmchm_s3:27}
\hat{\eta}&=&\cos\theta_{\rm{mix}}\eta_n+\sin\theta_{\rm{mix}}h_n ~,\nonumber \\
\hat{h}&=&-\sin\theta_{\rm{mix}}\eta_n+\cos\theta_{\rm{mix}}h_n ~.
\end{eqnarray}

\section{Level-splitting mechanism in the $\rm{SO(6)}/\rm{SO(5)}$ model}
\label{imp-fine}

In this section we study the improvement in the tension between the
light Higgs mass and the top-partner masses for a given $\Delta =
f^2/v_{ew}^2$ within the framework of the next-to-minimal model
introduced in the previous section.  We focus on the region of the
parameter space where this tension is released by the
\textit{level-splitting} mechanism that is operative when both $h$ and
$\eta$ develop vevs. We would confine ourselves to the region where
the vev of the singlet field $\eta$ is close to its natural value
$\chi \lesssim 1$, a choice which does not considerably worsen the
vev-tuning.  Thus $\Delta$ still notionally represents the minimal
vev-tuning in this model.

The potential given in Eq.~\eqref{nmchm_s3:16} can be minimised using
the expressions of Eq.~\eqref{nmchm_s3:17} as discussed in Appendix
\ref{appendix B}. This yields the mass matrix given in
Eq.~\eqref{nmchm_s3:20} which can be expressed in terms of the masses
of the two lightest resonances $Q_5$ and $Q_1.$ We will treat $\langle
h\rangle^2=\xi$ and $\langle\eta\rangle^2=\chi$ as free parameters in
our analysis. The decay constants of the vector-like fermionic
resonances $F^{t_L,t_R}$, defined in Appendix \ref{appendix B}, are
measures of compositeness of the left- and right-handed top
quarks. The compositeness fractions can be expressed as
\cite{Marzocca:2012zn}
\begin{equation}
\label{nmchm_s3:29}
\sin\phi_L\equiv\frac{\left|F^{t_L}\right|}{\sqrt{{m_{Q_5}^2}+\left|F^{t_L}\right|^2}}~,
\qquad
\sin\phi_R\equiv\frac{\left|F^{t_R}\right|}{\sqrt{{m_{Q_1}^2}+\left|F^{t_R}\right|^2}} ~.  
\end{equation}
We further  parametrise the  ratio of the decay constants by
\begin{equation}
\label{nmchm_s3:32}
\left|F^{t_L}\right|\equiv r\left|F^{t_R}\right|~.
\end{equation}  
The compositeness of $t_L$ is constrained to be smaller than $t_R$
from $Zb\overline{b}$ precision measurements \cite{Agashe:2005vg},
i.e. $r<1$. The expression for the physical top quark mass can be
extracted from Eq.~\eqref{nmchm_s3:15} as
\begin{eqnarray}
\label{nmchm_s3:30}
\label{nmchm_s3:31}
m_t^2&=&\frac{\left|M^t(0)\right|^2}{2}\xi\left( \chi
c_{2\theta}+(1-\xi)s_\theta^2\right) \nonumber \\ &=&
\frac{\left|F^{t_L}\right|^2\left|F^{t_R}\right|^2}{2m_{Q_1}^2m_{Q_5}^2}\bigg[m_{Q_1}^2
  +m_{Q_5}^2-2m_{Q_1}m_{Q_5}\cos\theta_{\rm{phase}}\bigg]\xi\left(
\chi c_{2\theta}+(1-\xi)s_\theta^2\right) ~,
\end{eqnarray} 
where $\theta_{\rm{phase}}$ corresponds to the phase associated with
the decay constants, which are in general be complex. Notice from the
above expression that for smaller values of $\theta_{\rm{phase}}$ the
top quark has to be more composite in order to generate $m_t\simeq 173$
GeV. 

The free parameters in the theory now reduce to $\theta, \xi, \chi, r,
\theta_{\rm{phase}}$ and the top-partner resonance masses $m_{Q_1},
m_{Q_5}$. To achieve our goal through the level-splitting mechanism we
need to ensure the following conditions:
\begin{enumerate}
 \item $m_{\eta_n\eta_n}^2>m_{h_nh_n}^2$ .
 \item $m_{h_n\eta_n}^2 \neq 0$ .
 \item The predominantly doublet state $(\hat{h})$ should have a mass $\sim 125$ GeV.
\end{enumerate}
\begin{figure}[h!]
\begin{subfigure}[]{0.52\textwidth}
\includegraphics[width=\linewidth]{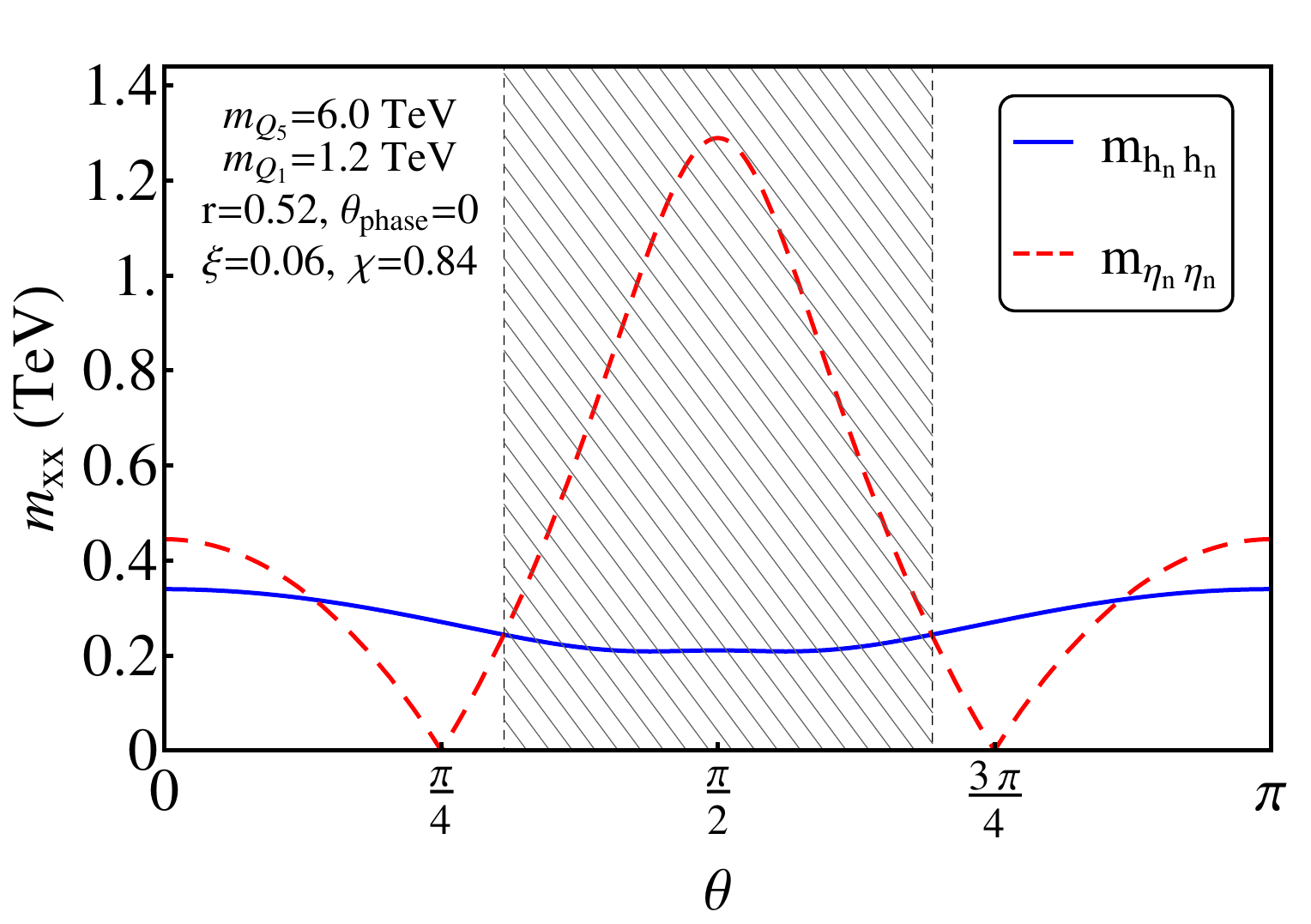}
\caption{}
\label{nmchm_s3:f2a}
\end{subfigure}%
\begin{subfigure}[]{0.52\textwidth}
\includegraphics[width=\linewidth]{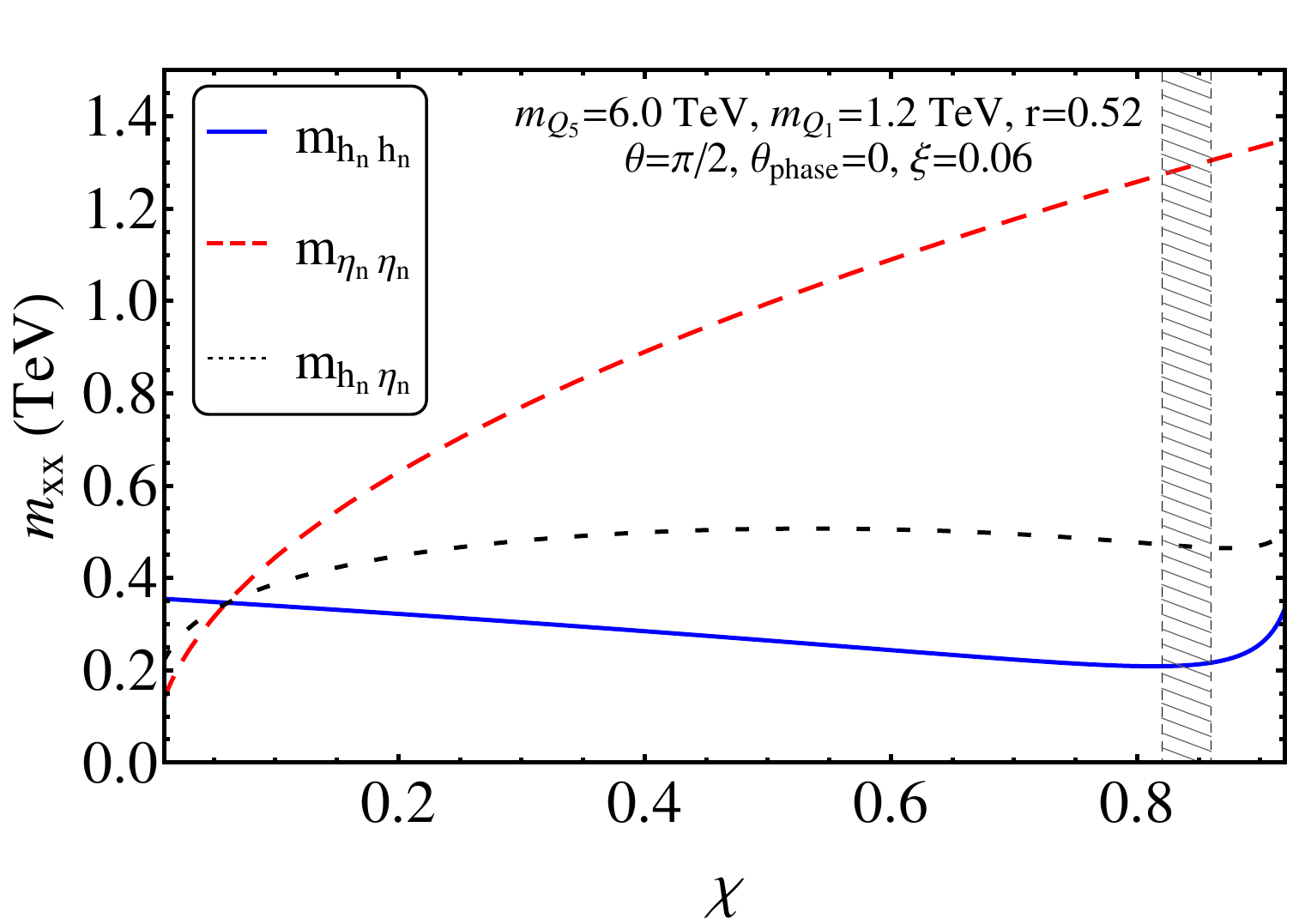}
\caption{}
\label{nmchm_s3:f2b}
\end{subfigure}
\begin{center}
\begin{subfigure}[]{0.52\textwidth}
\includegraphics[width=\linewidth]{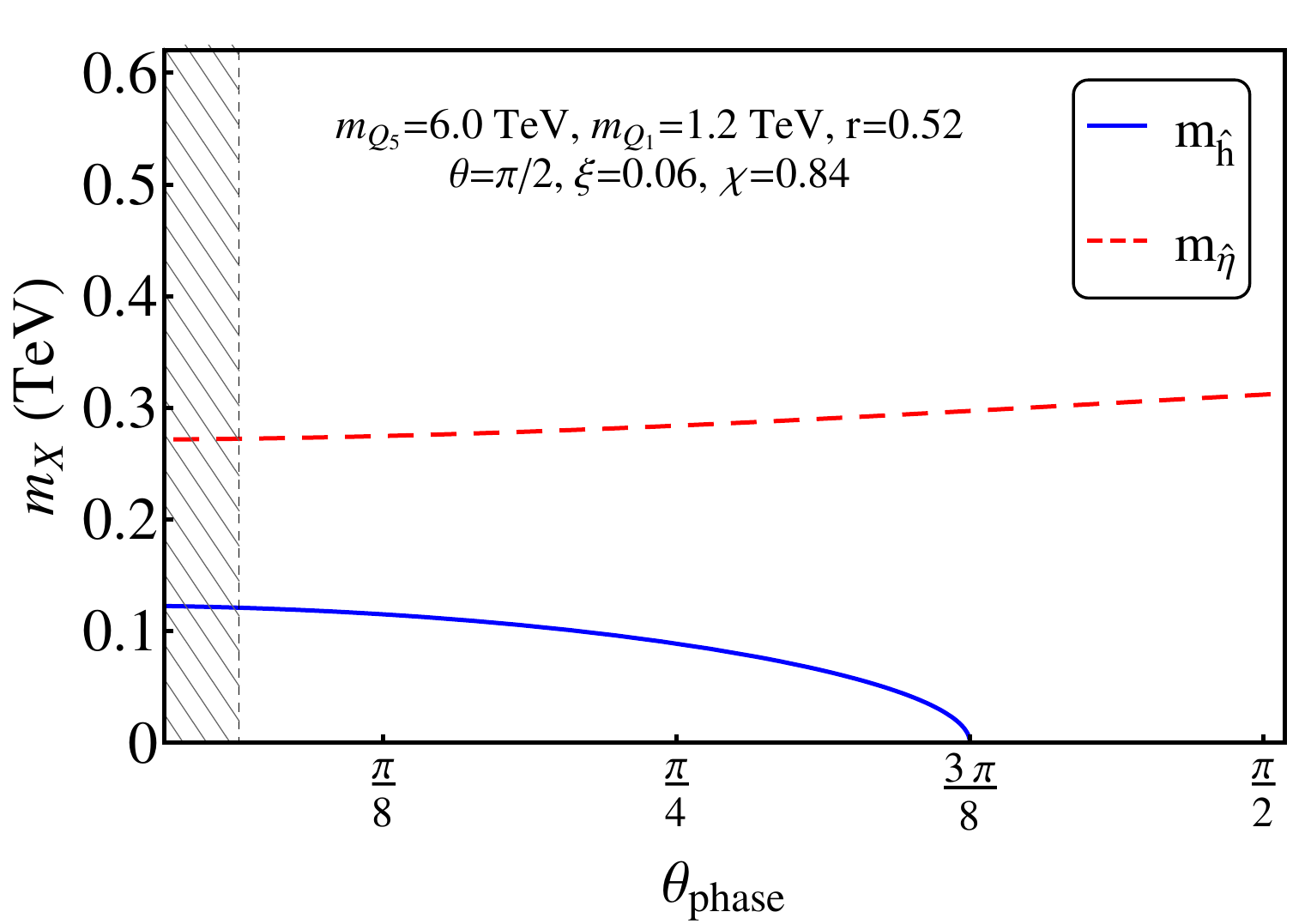}
\caption{}
\label{nmchm_s3:f2c}
\end{subfigure}
\end{center}
\caption{\small\it Different regions for the model parameters
  consistent with $m_{\hat{h}}=$125 GeV are displayed. In
  panel~\ref{nmchm_s3:f2a}, we show that the condition
  $m_{\eta_n\eta_n}^2>m_{h_nh_n}^2$ along with the requirement
  $\xi\ne0,~\chi\ne0$ are satisfied only in the shaded region around
  $\theta=\pi/2$. Panel~\ref{nmchm_s3:f2b} shows the variation of the
  mass matrix elements with the singlet vev $\chi$. The shaded strip
  corresponds to the range $m_{\hat{h}}=$[120-130] GeV.  In
  panel~\ref{nmchm_s3:f2c}, the shaded region is consistent with a
  light Higgs.}
\label{nmchm_s3:f2}
\end{figure}
 Na\"ively, one would expect that $m_{\eta_n\eta_n}^2>m_{h_nh_n}^2$
 would necessarily imply $\mu_2^2>\mu_1^2$. Note that $\mu_2^2$, being
 the parameter associated with the singlet field $\eta$, receives
 contribution only from loops involving $t_R$ (which is substantially
 composite as discussed later in the context of
 Fig.~\ref{nmchm_s3:f5b}), while $\mu_1^2$ receives contributions from
 both $t_L$ and $t_R$ loops in addition to a contribution from the gauge
 bosons. As a result, $\mu_1^2$ can even be larger than $\mu_2^2$ when
 all the contributions are combined. However, the interplay of
 $\lambda_1$, $\lambda_2$ and especially $\lambda_m$ through
 Eqs.~\eqref{nmchm_s3:17}, \eqref{nmchm_s3:21} and \eqref{nmchm_s3:22}
 can yield a substantial region where
 $m_{\eta_n\eta_n}^2>m_{h_nh_n}^2$ is satisfied. To demonstrate the
 choice of parameters we show the variation of the elements of the
 mass matrix given in Eq.~\eqref{nmchm_s3:20} as a function of the
 associated parameters in Fig.~\ref{nmchm_s3:f2}. As can be understood
 from the plot in Fig.~\ref{nmchm_s3:f2a},
 $m_{\eta_n\eta_n}^2>m_{h_nh_n}^2$ is satisfied when $\theta$ (as
 defined in Eq.~\eqref{nmchm_s3:14}) is near $0, \pi/2$ or $\pi.$
 However, the condition that both the doublet and the singlet receive
 vevs requires $\theta$ to be close to $\pi/2$ \footnote{While the
   doublet vev ensures EWSB, the singlet vev generates doublet-singlet
   mixing required for level-splitting. }. Note that $m_{h_nh_n}^2$
 and $m_{h_n\eta_n}^2$ are not quite sensitive to $\chi$ unlike
 $m_{\eta_n\eta_n}^2$. As demonstrated in Figs.~\ref{nmchm_s3:f2b} and
 \ref{nmchm_s3:f2c}, $m_{\hat{h}}\sim125$ GeV requires $\chi \gg \xi$
 and $\theta_{\rm{phase}}$ to be near zero.  Further the choice for
 $r$ is constrained by the measurements of the Higgs couplings at the
 LHC, as well as meeting the condition $\Pi^{t_L,t_R}_0\simeq 1$. The
 latter constraints prefer the region where $m_{Q_1}<m_{Q_5}$, which
 is the region explored in this paper.  Choosing these parameters
 admittedly results in additional tuning in the model. A quantitative
 estimate of this additional tuning asks for a statistical approach
 which is beyond the mandate of the present paper. In any case, we
 already obtain a considerable relaxation in the top-partner masses
 required to reproduce $m_{\hat{h}} = 125$ GeV for a given
 compositeness scale, as will be discussed below.

\begin{figure}[t]
\centering
\begin{subfigure}[t]{0.5\textwidth}
\centering
\includegraphics[width=\linewidth]{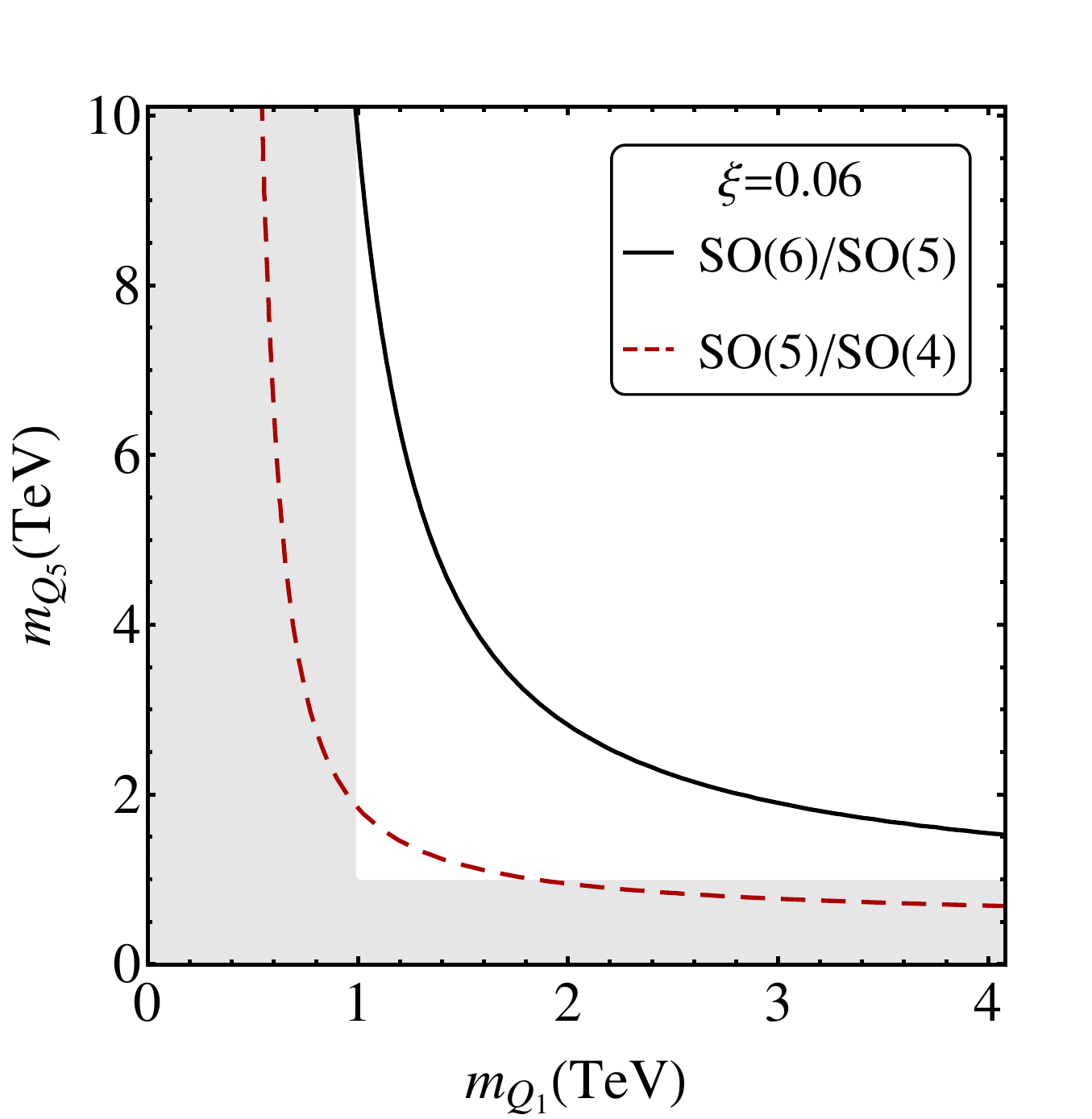}
\caption{}
\label{nmchm_s3:f1a}
\end{subfigure}%
~
\begin{subfigure}[t]{0.5\textwidth}
\centering
\includegraphics[width=\linewidth]{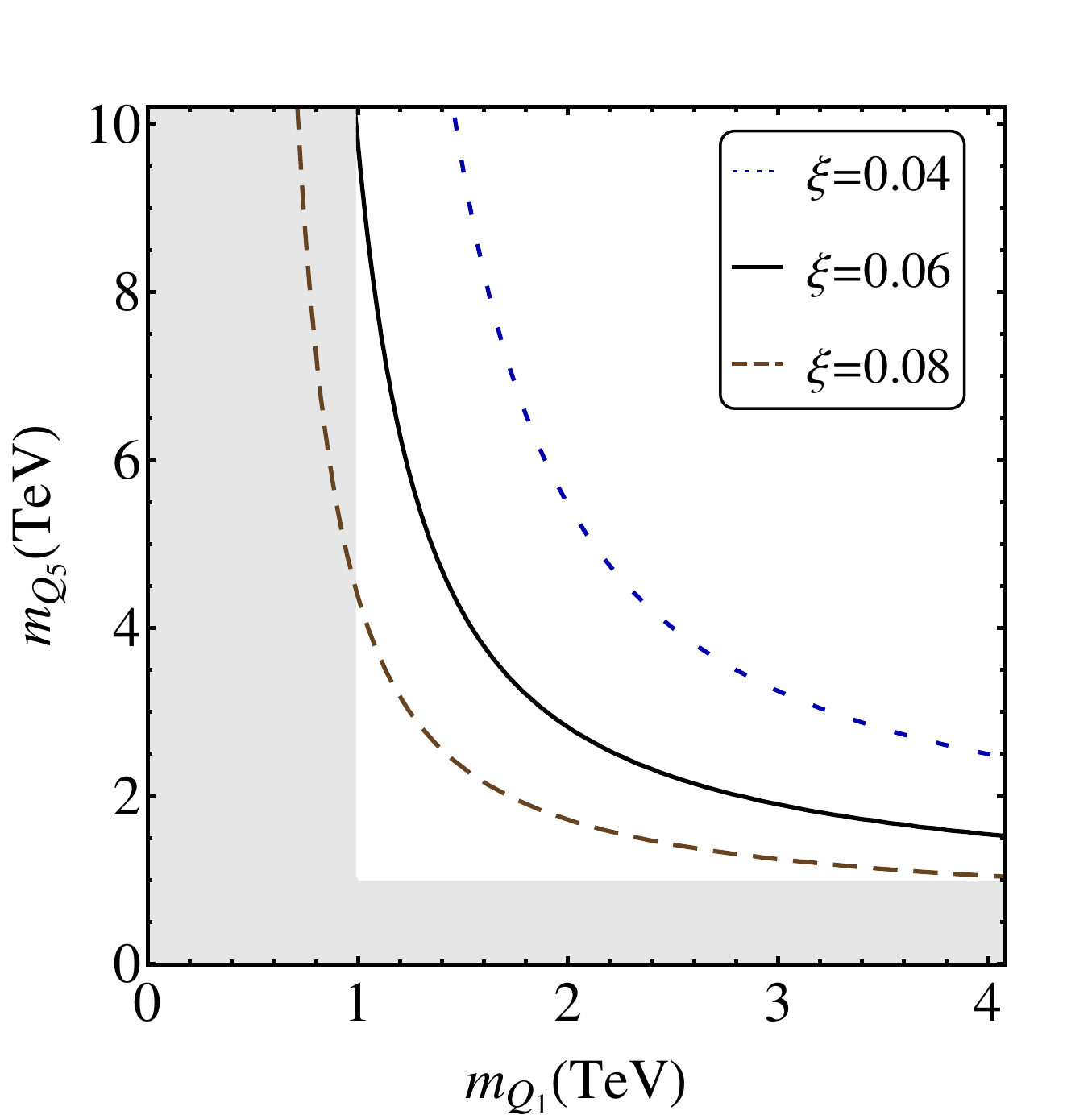}
\caption{}
\label{nmchm_s3:f1b}
\end{subfigure}
\caption{\small\it In panel~\ref{nmchm_s3:f1a} we show how
  level-splitting enables relaxation in the parameter space of masses
  of the top-partners for the same value of $\xi$. The plot
  corresponds to $\xi=0.06$ (i.e. $\Delta\simeq17$). The red dashed
  line shows the minimal model $m_{\hat{h}}=$125 GeV contour. Gray
  areas are already excluded by the LHC searches. The black contour
  refers to the next-to-minimal model. We have fixed
  $\theta=\pi/2,~\chi=0.84,~r=0.52.$ In panel~\ref{nmchm_s3:f1b} the
  three different contours are drawn for different values of $\xi$ in
  the next-to-minimal model, keeping $\theta$, $\chi$ and $r$ same as
  in panel~\ref{nmchm_s3:f1a}. For all cases, the doublet-singlet
  mixing is kept within $\theta_{\rm{mix}}<0.16$.}
\label{nmchm_s3:f1}
\end{figure} 

In Fig.~\ref{nmchm_s3:f1a} we demonstrate the quantitative impact of
level-splitting in the parameter space of the top-partner resonance
masses. Guided by the discussion above, we fix the other parameters
(see caption of Fig.~\ref{nmchm_s3:f1a}) to zoom into the region where
the relaxation of the top-partner masses is most pronounced.  The red
dashed line is the contour on which the Higgs mass is 125 GeV for the
minimal model for $\xi=0.06$.  Here we have mapped $m_{Q_5}
\rightarrow m_{Q_4}$ while comparing the contours for the minimal and
non-minimal models. For the same $\xi$, we find that the contour
shifts to heavier resonance masses away from the LHC direct search
limits. The level-repulsion mechanism is responsible for this
shift. The magnitude of this shift depends on the amount of
doublet-singlet mixing. Fig.~\ref{nmchm_s3:f1b} shows the Higgs mass
contours in the next-to-minimal setup for different choices of $\xi$.

\begin{figure}[t]
\centering
\begin{subfigure}[t]{0.5\textwidth}
\centering
\includegraphics[trim = 0mm 0mm 0mm 0mm, clip,width=77.5mm]{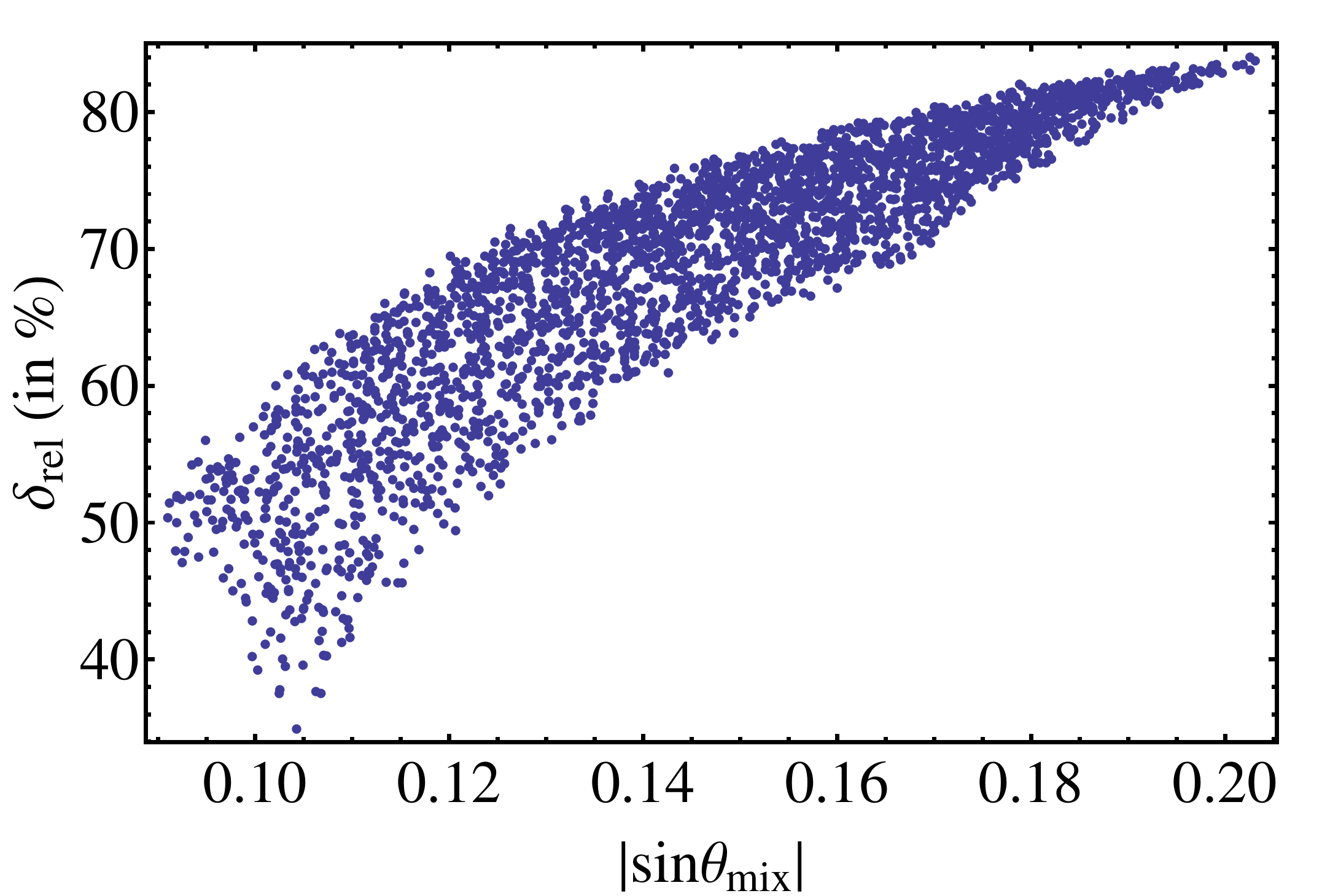}
\caption{}
\label{nmchm_s3:f3a}
\end{subfigure}%
~
\begin{subfigure}[t]{0.5\textwidth}
\centering
\includegraphics[trim = 0mm 0mm 0mm 0mm, clip,width=80mm]{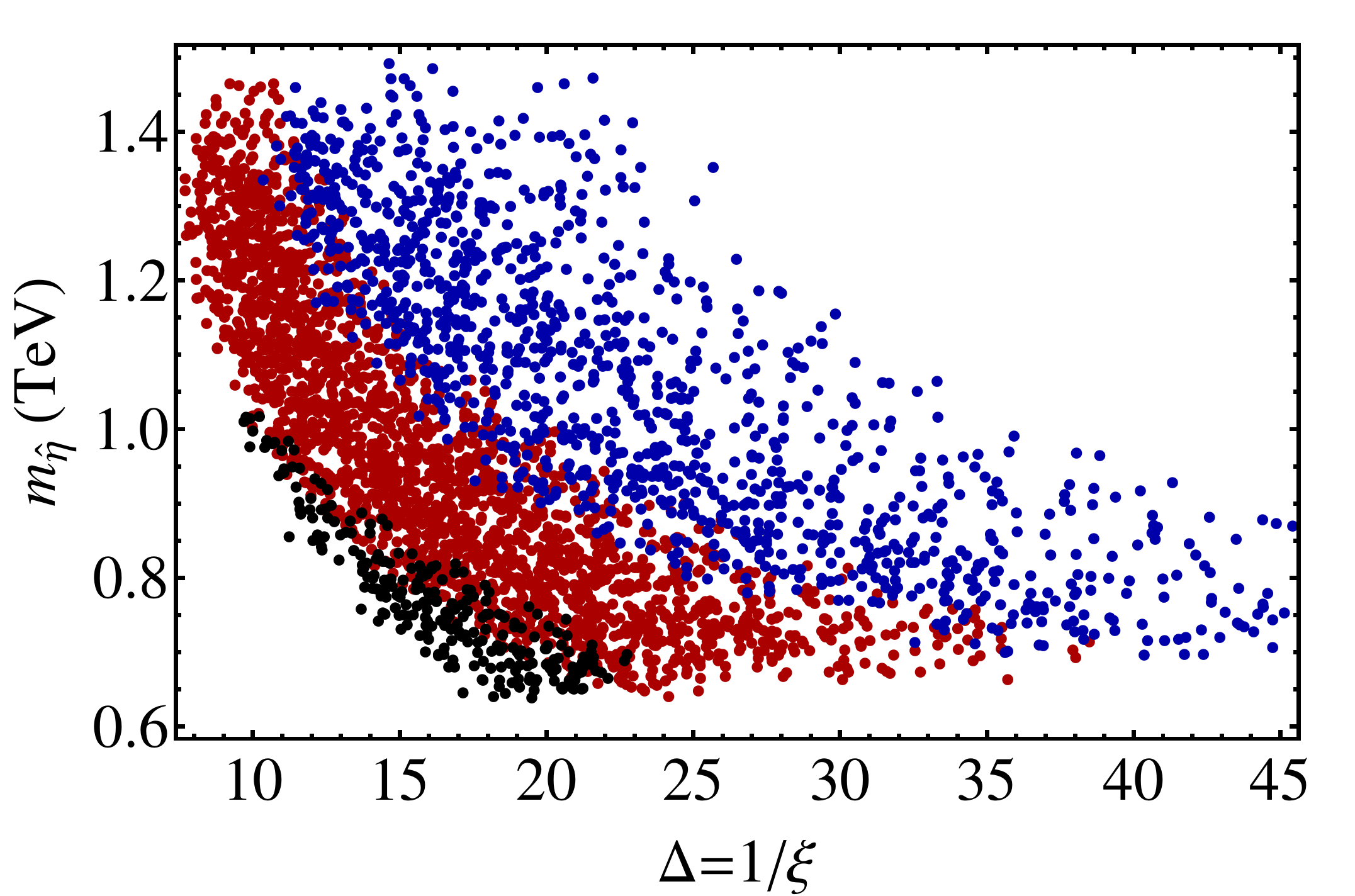}
\caption{}
\label{nmchm_s3:f3b}
\end{subfigure}
\caption{\small\it The plot on the left panel~\ref{nmchm_s3:f3a} shows
  the variation of percentage improvement in $\delta_{\rm{rel}}$,
  defined in Eq.~\eqref{nmchm_s3:33}, as a function of doublet-singlet
  mixing. The resonance masses $m_{Q_1}$ and $m_{Q_5}$ are taken in
  the range between [1.0-1.5] TeV and [5.0-6.0] TeV, respectively and
  $\chi$ is varied within the range [0.4-0.9] keeping $r=0.52$. In
  panel~\ref{nmchm_s3:f3b} we have plotted the mass of the heavier
  singlet-like eigenstate ($m_{\hat{\eta}}$) as a function of
  $\Delta$. Blue points correspond to $\theta_{\rm{mix}}<0.15$, red to
  $0.15<\theta_{\rm{mix}}<0.2$ and black points to
  $0.2<\theta_{\rm{mix}}<0.3$. Here, $m_{Q_1}$ and $m_{Q_5}$ are
  varied in the ranges [1.0-1.2] TeV and [5.5-6.0] TeV,
  respectively. The parameters $\chi$ and r are varied in the ranges
  [0.75-0.90] and [0.45-0.90], respectively.}
\label{nmchm_s3:f3}
\end{figure}

The percentage improvement in $\Delta$ in this model compared to the
minimal model is estimated as
\begin{equation}
\label{nmchm_s3:33}
\delta_{\rm{rel}}=\frac{\left|\Delta_{\rm{SO(5)}/\rm{SO(4)}}-\Delta_{\rm{SO(6)}/\rm{SO(5)}}\right|}{\Delta_{\rm{SO(5)}/\rm{SO(4)}}}\times
100\% ~,
\end{equation} 
where for standardization we have kept the masses of the lowest two
top-partners identical in both $\rm{SO(5)}/\rm{SO(4)}$ and
$\rm{SO(6)}/\rm{SO(5)}$ setups. In Fig.~\ref{nmchm_s3:f3a} we show the
improvement of fine-tuning as a function of the doublet-singlet mixing
angle $\theta_{\rm{mix}}$, defined in Eq.~\eqref{nmchm_s3:26}.  In
Fig.~\ref{nmchm_s3:f3b}, we plot the mass of the dominantly singlet
state with $\Delta$. It is evident from Fig.~\ref{nmchm_s3:f3b} that a
smaller $\Delta$ can be obtained at the expense of increasing
$m_{\hat{\eta}}$. Also, for the same $\Delta $ larger mixing results
in smaller $m_{\hat{\eta}}$, as evident from Eq.~\eqref{nmchm_s3:26}.

\section{Phenomenological consequences}
\label{pheno}
In this section we discuss some of the phenomenological consequences
of this setup. The doublet-singlet scalar mixing ensures that the
Higgs couplings to massive gauge bosons are further suppressed
compared to the minimal model. This can provide an avenue to probe and
constrain this framework.  In the region of parameter space of
interest (namely, $m_{\eta_n\eta_n}^2>m_{h_nh_n}^2$), the top quark
(mainly the right-handed component) turns out to be substantially
composite.  And finally, the singlet-like state $\hat{\eta}$ is
unstable and its decay signatures would leave tangible imprint in
colliders.

\begin{figure}[t]
\centering
\includegraphics[trim = 0mm 0mm 0mm 0mm, clip,width=120mm]{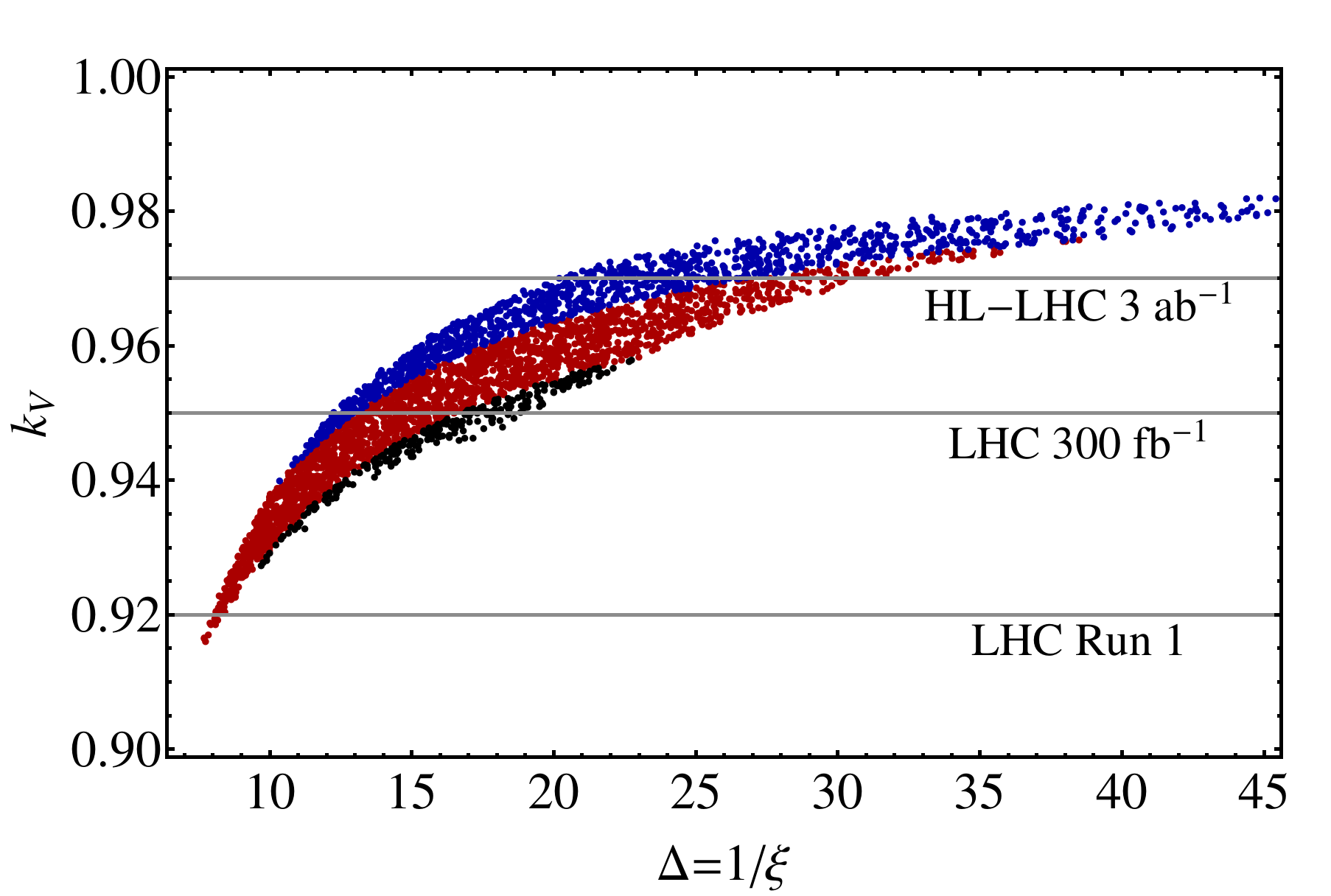}
\caption{\small\it The variation of $k_V$, the modification factor in
  $hVV$ coupling defined in Eq.~\eqref{nmchm_s3:34}, with the
  parameter $\Delta$ is shown here. The horizontal gray lines
  represent the $1\sigma$ present \cite{Khachatryan:2016vau} and
  anticipated \cite{CMS:2013xfa, ATL-PHYS-PUB-2014-016} LHC limits
  with different luminosities.  We have varied $m_{Q_1}$ and $m_{Q_5}$
  in the ranges [1.0-1.2] TeV and [5.5-6.0] TeV, respectively. The
  parameters $\chi$ and r are varied in the ranges [0.75-0.90] and
  [0.45-0.90], respectively. For the colours of different scattered
  points, see caption of Fig.~\ref{nmchm_s3:f3b} (essentially
  $\theta_{\rm{mix}}$ decreases as we go up).}
\label{nmchm_s3:f4}
\end{figure}

Constraints on the doublet-singlet mixing come mainly from the Higgs
couplings with the gauge bosons. Like in the minimal model, the Higgs
couplings to the massive gauge bosons are already suppressed by a
factor of $\sqrt{1-\xi}$ with respect to the SM value. On top of that,
doublet-singlet mixing induces an additional suppression quantified by
$\theta_{\rm{mix}}$. This suppression can be parametrised as
\begin{equation}
\label{nmchm_s3:34}
k_V=\frac{g_{\hat{h}VV}}{g_{\hat{h}VV}^{SM}}=\cos\theta_{\rm{mix}}\sqrt{1-\xi} ~.
\end{equation}
Within the minimal model, a lower bound $f \gtrsim 700$ GeV
\cite{Falkowski:2013dza} was obtained from Higgs physics. This is a
somewhat conservative estimate compared to the limit
\cite{Barbieri:2007bh} obtained from electroweak precision tests which
involve uncertainties stemming from some incalculable UV dynamics. In
the next-to-minimal model, the extra suppression strengthens the above
limit on $f$. In fact, we have estimated that with $\theta_{\rm{mix}}
= 0.2$, the lower bound increases to $f\gtrsim 850$ GeV. For all the
parameter choices that have gone into Fig.~\ref{nmchm_s3:f1}, the
doublet-singlet mixing is always kept below $\theta_{\rm{mix}} =
0.16$, and $f>850$ GeV (i.e. $\xi<0.084$).  In Fig.~\ref{nmchm_s3:f4}
we present the deviation of the Higgs couplings to massive gauge
bosons, defined in Eq.~\eqref{nmchm_s3:34}, with the parameter
$\Delta$. The plot shows that even a moderate $\Delta = 10$ is well
within the LHC tolerance limit \cite{Khachatryan:2016vau}.  However,
future Higgs branching ratio measurements with higher precision would
challenge such tolerance \cite{CMS:2013xfa,
  ATL-PHYS-PUB-2014-016}. Additionally, mixing between CP-even and
CP-odd states would also have consequences testable in future
measurements. It is worth noting that the physical Higgs couplings
with the gauge bosons are always suppressed with respect to both the
minimal composite model as well as the SM. On the other hand, the
coupling of the physical Higgs to the top quark is not necessarily
suppressed which may lead to interesting phenomenological
consequences.

\begin{figure}[t]
\centering
\begin{subfigure}[t]{0.5\textwidth}
\centering
\includegraphics[width=\linewidth]{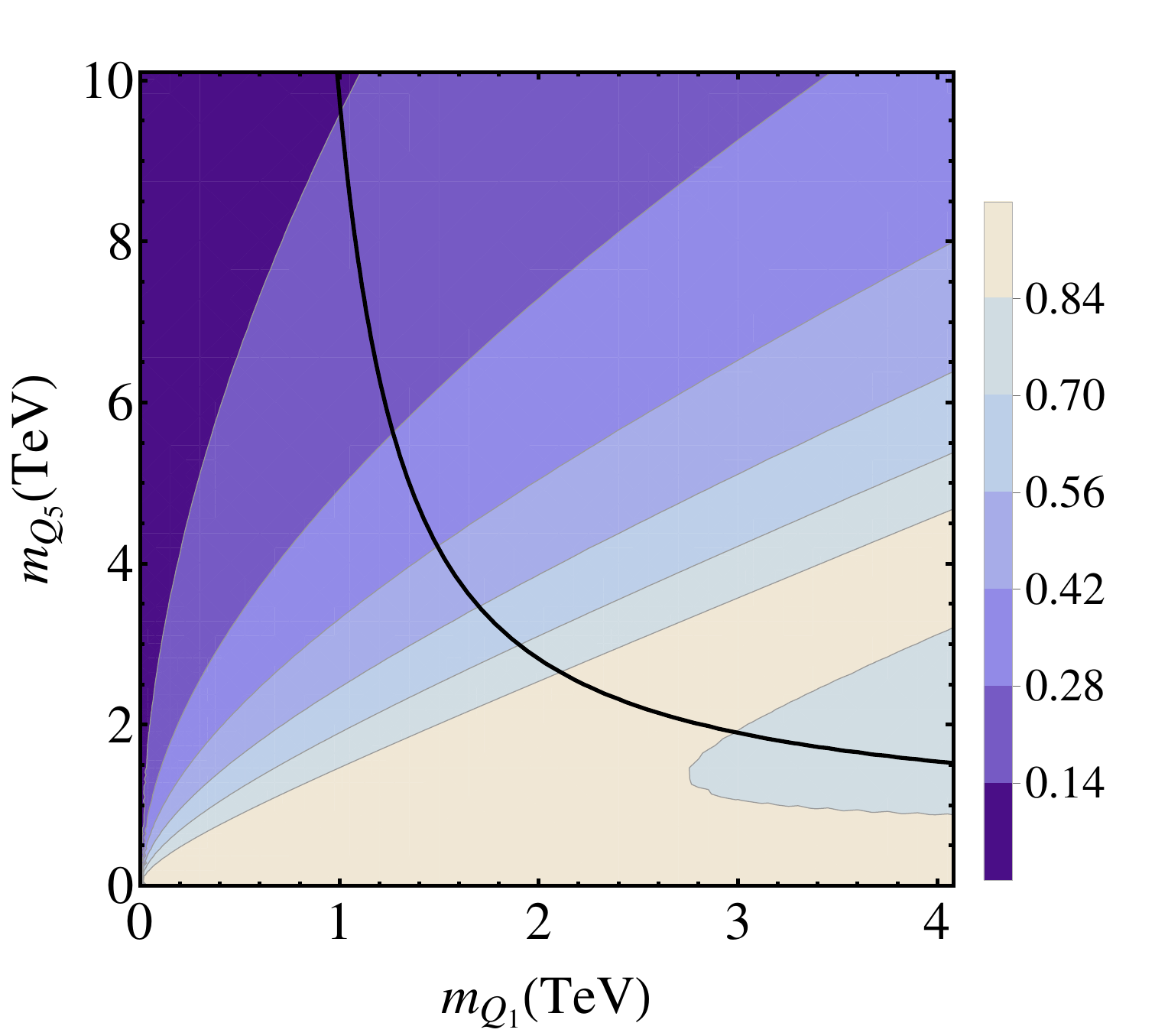}
\caption{}
\label{nmchm_s3:f5a}
\end{subfigure}%
~
\begin{subfigure}[t]{0.5\textwidth}
\centering
\includegraphics[width=\linewidth]{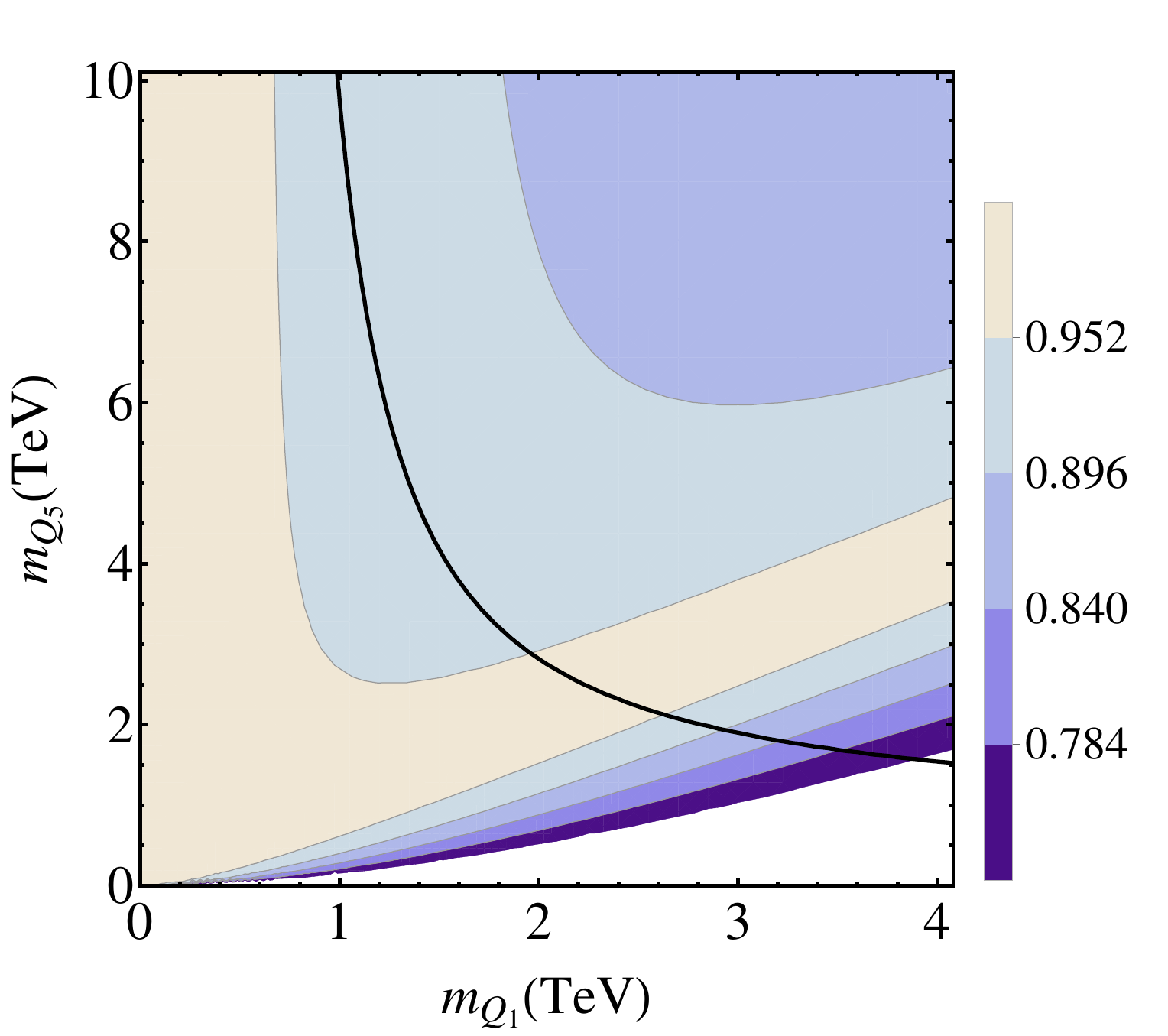}
\caption{}
\label{nmchm_s3:f5b}
\end{subfigure}
\caption{\small\it The background shades representing the compositeness
  fractions of $t_L$ and $t_R$, defined in Eq.~\eqref{nmchm_s3:29}, are shown in Figs.~\ref{nmchm_s3:f5a}
  and \ref{nmchm_s3:f5b}, respectively. The value of $\xi$ is fixed
  at 0.06 for both the plots. The black line refers to
  $m_{\hat{h}}=$125 GeV, as in Fig.~\ref{nmchm_s3:f1a}.}
\label{nmchm_s3:f5}
\end{figure}

In Figs.~\ref{nmchm_s3:f5a} and \ref{nmchm_s3:f5b}, we have plotted
the compositeness fraction of $t_L$ and $t_R$, defined in
Eq.~\eqref{nmchm_s3:29}, in the parameter space of the top-partner
masses. Note that $t_L$ turns out to be relatively elementary
\cite{Agashe:2005vg}\footnote{However, in the region where $m_{Q_1}
  \sim m_{Q_5}$, $t_L$ appears to be composite. This is a
  calculational artifact of forcing $r=|F^{t_L}|/|F^{t_R}|$ to fixed
  values for simplified presentation in the plot and simultaneously
  determining $m_t$ from Eq.~\eqref{nmchm_s3:31} using
  $\theta_{\rm{phase}}=0$. In fact, $t_L$ can be consistently kept
  mostly elementary by choosing appropriate values of $r$.}, while
$t_R$ is mostly composite in a large region of viable parameter space.
Suggested studies to probe the compositeness of $t_R$
\cite{Gounaris:2016bir, Renard:2017qjn} would provide another handle
to constrain and explore this mechanism.

Finally, we briefly comment on the phenomenology of the additional
singlet-like state $\hat{\eta}$, whose detailed collider phenomenology
has been studied in \cite{Arbey:2015exa,Niehoff:2016zso}. Since it has
a small doublet component depending on the mixing parameter
$\theta_{\rm{mix}},$ it has nontrivial coupling to the gauge
bosons. It also has a nontrivial coupling to the third generation
quarks.  If the mass of $\hat{\eta}$ is within the range of the LHC it
can be produced in the same way as the 125 GeV Higgs has been produced
with the maximum contribution coming from gluon fusion. For large
$\hat{\eta}$ mass the production will be suppressed even for sizable
mixing. For $m_{\hat{\eta}}>2m_{\hat{h}}$ and $m_{\hat{\eta}}>2m_{t}$,
novel channels like $\hat{\eta}\rightarrow\hat{h}\hat{h}$ and
$\hat{\eta}\rightarrow t\overline{t}$ would open up. As shown in
\cite{Niehoff:2016zso}, for the choice of $m_{\hat{\eta}} = 1$ TeV,
the production cross section of $\hat\eta$ times its branching ratio
into $\hat{h}\hat{h}$ channel in LHC (13 TeV) lies in the range $(0.01
- 0.1)$ pb, and the same for $t\bar{t}$ channel is two orders of
magnitude smaller. While the CMS and ATLAS exclusion limits on the
same quantity for the $\hat{h}\hat{h}$ final states hovers around the
predicted upper limit, the experimental exclusion limits for the
$t\bar{t}$ channel currently lie substantially above the predicted
numbers. As regards the production cross section times the branching
ratio in the diphoton channel, the theory prediction for
$m_{\hat{\eta}} = 1$ TeV is in the range $(10^{-5} - 10^{-4})$ pb,
while the CMS and ATLAS sensitivities lie two orders above.

\section{Conclusions}
\label{s4}
In this paper we have explored the scalar sector of the
next-to-minimal composite Higgs model, having a coset
$\rm{SO(6)}/\rm{SO(5)}$, where the set of pNGB states contains an
additional SM singlet compared to the minimal $\rm{SO(5)}/\rm{SO(4)}$
model. We demonstrate that a relaxation in the top-partner mass for a
given compositeness scale and the measured Higgs mass is achieved in
the next-to-minimal model employing the $\textit{level-splitting}$
mechanism.  This is operative in a generic minima of the potential
where both the doublet and the singlet scalars receive vevs leading to
a non-trivial doublet-singlet mixing. Consequently, the relatively
lighter doublet-like state becomes even more lighter, to be identified
with the observed $125$ GeV Higgs boson, while the mass of the
singlet-like state is further jacked up.  We have performed an
extensive scan over the parameter space where the theory is in
agreement with all phenomenological constraints. We observe a
substantial improvement in the minimal vev-tuning parameter
$\Delta$. This improvement can be phrased through the modification of
Eq.~\eqref{nmchm_s1:1}, using Eq.~\eqref{nmchm_s3:25}, as follows,
\begin{equation}
\label{nmchm_conc:1}
m_{\hat{h}}^2\sim \frac{N_c}{\pi^2}y_t^2
\frac{m_Q^2}{\Delta}-m_{h_n\eta_n}^2\tan\theta_{\rm{mix}}~,
\end{equation}
where the second term in the right-hand side is necessarily
positive. Eq.~\eqref{nmchm_conc:1} thus allows for a larger value of
$m_Q$ compared to what is admissible by Eq.~\eqref{nmchm_s1:1} for the
same choice of $\Delta$.  Admittedly, the enhanced naturalness came at
the expense of adding an extra singlet scalar which descended from the
enlarged coset of the present scenario compared to the minimal model.

There are three main phenomenological consequences of this
setup. First, there would be a larger deviation in the Higgs couplings
to the $W$ and $Z$ bosons compared to minimal case. The Higgs coupling
to the top quark would be modified too. The deviations are still
within the LHC limit but would start getting constrained with more
precise measurements of the Higgs branching ratios.  Second, the
pronounced compositeness of the right-handed top quark that dominates
the parameter space of our interest is worth noting. Searches for
$t_R$-compositeness in colliders can be a tool to probe such a
mechanism.  Finally, the phenomenology of the singlet-like scalar
$\hat{\eta}$ would resemble that of the observed Higgs boson
$\it{modulo}$ a significant suppression in its couplings.

\section*{Acknowledgments}

We thank Tomasz Dutka, Tony Gherghetta and Raymond Volkas for
collaboration during the early stages of the project. Discussions with
Tony Gherghetta, Aldo Deandrea, Ben Gripaios and Giacomo Cacciapaglia
are gratefully acknowledged. We also thank the participants of the
Indo-French LIA THEP and CEFIPRA INFRE-HEPNET Kick-off meeting at
IISc, Bangalore (2016), for raising concerns on the issue of
fine-tuning in composite Higgs models in general. AB acknowledges
financial support from Department of Atomic Energy, Government of
India. GB acknowledges support of the J.C.\ Bose National Fellowship
from the Department of Science and Technology, Government of India
(SERB Grant No.\ SB/S2/JCB-062/2016). TSR is partially supported by
the Department of Science and Technology, Government of India, under
the Grant Agreement number IFA13-PH-74 (INSPIRE Faculty Award).

\appendix

\section{Effective potential in $\rm SO(5)/SO(4)$ model}
\label{appendix minimal model}

The Lagrangian for top-quark (in fundamental representation of $\rm SO(5)$) is given by
\begin{eqnarray}
\label{nmchm_ap min:1}
\mathcal{L}=\overline{t}_L\slashed{p}\left[
  \Pi^{t_L}_0+\frac{\tilde{\Pi}^{t_L}_1}{2}h^2\right]t_L+\overline{t}_R\slashed{p}\left[
  \Pi^{t_R}_0+\tilde{\Pi}^{t_R}_1(1-h^2)\right]t_R+\overline{t}_L\left[\frac{M^{t}}{\sqrt{2}}h\sqrt{1-h^2}\right]t_R
+\rm{h.c.}
\end{eqnarray}
The top-quark contribution to one-loop C-W potential, obtained using the above Lagrangian can be expressed as
\begin{equation}
\label{nmchm_ap min:2}
V_{\rm{top}}(h)= -\frac{\mu_t^2}{2}h^2+\frac{\lambda_t}{4}h^4~. 
\end{equation} 
The parameters $\mu_t^2$ and $\lambda_t$  are calculated by integrating over the structure functions. We use the Weinberg sum rules to model the structure functions and introduce minimal set of resonances required to saturate the integrals making them finite. Following two conditions on each of the form factors are needed for the convergence of $\lambda_t$:
\begin{eqnarray}
\label{nmchm_ap min:3}
\lim\limits_{q_E^2\rightarrow 0} q_E^n\frac{\tilde{\Pi}^{t_L,t_R}_1}{\Pi^{t_L,t_R}_0}=0~,~(n=0,2) ~,
\end{eqnarray}
whereas, to make $\mu_t^2$ calculable, one extra condition is required, which gives rise to one more sum rule: 
\begin{eqnarray}
\label{nmchm_ap min:4}
\lim\limits_{q_E^2\rightarrow 0} \left(\frac{\tilde{\Pi}^{t_L}_1}{2\Pi^{t_L}_0}-\frac{\tilde{\Pi}^{t_R}_1}{\Pi^{t_R}_0}\right)=0 ~.
\end{eqnarray}
The parameters in the potential can  now be calculated as
\begin{eqnarray}
\nonumber
\frac{\mu_t^2}{2} = \frac{\lambda_t}{4} &=& 2N_c\int \frac{d^4q_E}{(2\pi)^4} \left[\frac{1}{8}\left(\frac{\tilde{\Pi}^{t_L}_1}{\Pi^{t_L}_0}\right)^2+\frac{1}{2}\left(\frac{\tilde{\Pi}^{t_R}_1}{\Pi^{t_R}_0}\right)^2+ \frac{|M^{t}|^2}{2q_E^2\Pi^{t_L}_0\Pi^{t_R}_0}\right]\\
\label{nmchm_ap min:5}
&=&\frac{N_c}{8\pi^2}\frac{m_t^2m_{Q_1}^2m_{Q_4}^2}{m_{Q_1}^2-m_{Q_4}^2}\log\left(\frac{m_{Q_1}^2}{m_{Q_4}^2}\right)\frac{1}{\xi(1-\xi)} ~,
\end{eqnarray}
where $m_t$ is the top quark mass and $m_{Q_1}$ and $m_{Q_4}$ represent the masses corresponding to the two lightest top-partners transforming as singlet and quadruplet under  $\rm{SO(4)}$, respectively. It follows from  Eq. \eqref{nmchm_ap min:5} that, the top-quark contribution alone gives $\xi=0.5$. The gauge sector contribution on the other hand, enabling a cancellation with fermion contribution, can effectively reduce $\xi$. Considering both gauge and top contributions, the total potential can be written as
\begin{equation}
\label{nmchm_ap min:6}
V_{\rm{eff}}(h)= -\frac{\mu^2}{2}h^2+\frac{\lambda}{4}h^4= -\frac{1}{2}(\mu_t^2-\mu_g^2) h^2+\frac{1}{4}(\lambda_t-\lambda_g)h^4~,
\end{equation} 
where\cite{Pomarol:2012qf}
\begin{eqnarray}
\label{nmchm_ap min:7}
\frac{\mu_g^2}{2}=\frac{9}{2}\int\frac{d^4q_E}{(2\mathnormal{\pi})^4} \frac{\Pi_1(-q_E^2)}{4\Pi_0(-q_E^2)}=\frac{9g^2\mathnormal{f}^2m_{\rho}^2m_a^2}{128\pi^2(m_a^2-m_{\rho}^2)}\log\left(\frac{m_a^2}{m_{\rho}^2}\right)~,\\
\nonumber
\frac{\lambda_g}{4}=-\frac{9}{2}\int\frac{d^4q_E}{(2\mathnormal{\pi})^4}\frac{\Pi_1(-q_E^2)^2}{32\Pi_0(-q_E^2)^2}=-\frac{9g^4\mathnormal{f}^4}{1024\pi^2}\biggl[\log\left(\frac{m_am_\rho}{M_W^2}\right)-\frac{(m_a^4+m_\rho^4)}{(m_a^2-m_\rho^2)^2}\\
\label{nmchm_ap min:8}
-\frac{(m_a^2+m_\rho^2)(m_a^4-4m_a^2m_{\rho}^2+m_{\rho}^4)}{2(m_a^2-m_{\rho}^2)^3}\log\left(\frac{m_a^2}{m_{\rho}^2}\right)\biggr]~,
\end{eqnarray}
where $m_a, m_\rho$ are masses of the vector resonances in the strong sector. These are also expected to be light to account for the perturbative unitarity of the theory \cite{Bellazzini:2012tv}. It is crucial to note that, $\mu_g^2$ is always positive. Clearly, to drive EWSB, $\mu_g^2<\mu_t^2$ is a requirement. The numerical impact of $\lambda_g$ is small.

\section{$\rm{SO(6)}$ Algebra}
\label{appendix A}

We have used the following convention  for the generators of $\rm{SO(6)}$: 
\begin{eqnarray}
\nonumber
T^A=\{T^{\alpha},\hat{T}^{\hat{\alpha}} \}=\{ T^a_L,T^a_R,T^{\hat{a}},\hat{T}^{\hat{\alpha}} \};~~~~~~~~~~~~~~~
\\
\nonumber
(A=1,...,15;~\alpha=1,...,10;~\hat{\alpha}=1,...,5;~a=1,2,3;~\hat{a}=1,2,3,4)~,
\end{eqnarray}
where $\alpha=\{a,\hat{a}\}$ denotes the unbroken indices and $\hat{\alpha}$ denotes the broken indices for $\rm{SO(6)}\rightarrow\rm{SO(5)}$. The explicit expressions of the generators in fundamental representation of $\rm{SO(6)}$ are as follows:
\begin{eqnarray}
\label{nmchm_aC:1}
(T^a_{L,R})_{ij}&=&-\frac{i}{2}\left[\frac{1}{2}\epsilon^{abc}(\delta^b_i\delta^c_j-\delta^b_j\delta^c_i)\pm (\delta^a_i\delta^4_j-\delta^a_j\delta^4_i)\right]~,\\
\label{nmchm_aC:2}
(T^{\hat{a}})_{ij}&=&-\frac{i}{\sqrt{2}}(\delta^{\hat{a}}_i\delta^5_j-\delta^{\hat{a}}_j\delta^5_i)~,\\
\label{nmchm_aC:3}
(\hat{T}^{\hat{\alpha}})_{ij}&=&-\frac{i}{\sqrt{2}}(\delta^{\hat{\alpha}}_i\delta^6_j-\delta^{\hat{\alpha}}_j\delta^6_i)~,
\end{eqnarray}
where $i,j$ run from 1 to 6.

\section{Effective potential from the top sector for $\rm{SO(6)}/\rm{SO(5)}$ model}
\label{appendix B}

With the elementary top quark embeddings  given in Eqs.~\eqref{nmchm_s3:13} and \eqref{nmchm_s3:14}, the effective Lagrangian involving the top quark can be constructed by taking group theoretic invariants as follows, 
\begin{eqnarray}
\nonumber
\mathcal{L}=\Pi^{t_L}_0(p)\overline{t}_L\slashed{p}t_L+ \tilde{\Pi}^{t_L}_1(p)(\overline{Q}_L\Sigma)\slashed{p}(\Sigma^T Q_L)+\Pi^{t_R}_0(p)\overline{t}_R\slashed{p}t_R+ \tilde{\Pi}^{t_R}_1(p)(\overline{T}_R\Sigma)\slashed{p}(\Sigma^T T_R)\\
\label{nmchm_B.3:1}
+M^{t}(p)(\overline{Q}_L\Sigma)(\Sigma^T T_R)+\rm{h.c.}~,~
\end{eqnarray}
where $\Sigma$ is given in Eq.~\eqref{nmchm_s3:4}, and the momentum
dependent form factors encode the details of the dynamics of composite
resonances.  Substituting explicit forms of $Q_L,~T_R$ and $\Sigma$,
as given in Eqs.~\eqref{nmchm_s3:13}, \eqref{nmchm_s3:14} and
\eqref{nmchm_s3:4} respectively, the effective Lagrangian takes the
following form:
\begin{eqnarray}
\nonumber
\mathcal{L}=\overline{t}_L\slashed{p}\left[ \Pi^{t_L}_0+\frac{\tilde{\Pi}^{t_L}_1}{2}h^2\right]t_L+\overline{t}_R\slashed{p}\left[ \Pi^{t_R}_0+\tilde{\Pi}^{t_R}_1\left(
\label{nmchm_B.3:2} 
c_{2\theta}\eta^2+s_\theta^2(1-h^2)\right)\right]t_R\\
+\overline{t}_L\left[\frac{M^{t}}{\sqrt{2}}h\left( ic_\theta\eta+s_\theta\sqrt{1-h^2-\eta^2}\right)\right]t_R +\rm{h.c.}
\end{eqnarray}
This leads to the following one-loop effective Coleman-Weinberg potential:
\begin{eqnarray}
\nonumber
V_{\rm{eff}}(h,\eta)=-2N_c\int \frac{d^4q_E}{(2\pi)^4} &\left[ \log\left(1+\frac{\tilde{\Pi}^{t_L}_1}{2\Pi^{t_L}_0}h^2\right)+\log\left(1+\frac{\tilde{\Pi}^{t_R}_1}{\Pi^{t_R}_0}( c_{2\theta}\eta^2+s_\theta^2(1-h^2))\right)\right.\\
\label{nmchm_B.3:3}
&+\left.\log\left(1+\frac{|M^{t}|^2}{2q_E^2\Pi^{t_L}_0\Pi^{t_R}_0}h^2( c_{2\theta}\eta^2+s_\theta^2(1-h^2))\right)\right] ~,
\end{eqnarray}
where $N_c$ is the number of QCD colours of the fermions ($N_c=3$), and $q_E$ denotes the Euclidean momenta. The form factors $\Pi_0,~\tilde{\Pi}_1,~M^t$ can be split keeping in mind $\rm{SO(5)}$ representations as follows,
\begin{eqnarray}
\label{nmchm_B.3:4}
\Pi^{t_L,t_R}_0=1+\Pi^{t_L,t_R}_{Q_ 5}~,
\qquad
\tilde{\Pi}^{t_L,t_R}_1=\Pi^{t_L,t_R}_{Q_1}-\Pi^{t_L,t_R}_{Q_5}~,
\qquad
M^{t}=M^{t}_{Q_1}-M^{t}_{Q_5} ~.
\end{eqnarray}
$Q_1$ and $Q_5$ represent resonances transforming as singlet and five-plet under $\rm{SO(5)}$, respectively. Now, in view of the results obtained from large-N formalism, form factors encoding strong dynamics can be written as a sum of contributions from infinite number of resonance particles with increasing mass as
\begin{eqnarray}
\label{nmchm_B.3:5}
\Pi^{t_L,t_R}_{Q_5}&=&\sum_{n}^{}\frac{\left|F^{t_L,t_R}_{Q_5^{(n)}}\right|^2}{q_E^2+m_{Q_5^{(n)}}^2}~,
\qquad
\Pi^{t_L,t_R}_{Q_1}=\sum_{n}^{}\frac{\left|F^{t_L,t_R}_{Q_1^{(n)}}\right|^2}{q_E^2+m_{Q_1^{(n)}}^2}~,
\\
\label{nmchm_B.3:6}
M^{t}_{Q_5}&=&\sum_{n}^{}\frac{F^{t_L}_{Q_5^{(n)}}F^{*t_R}_{Q_5^{(n)}}m_{Q_5^{(n)}}}{q_E^2+m_{Q_5^{(n)}}^2}~,
\qquad
M^{t}_{Q_1}=\sum_{n}^{}\frac{F^{t_L}_{Q_1^{(n)}}F^{*t_R}_{Q_1^{(n)}}m_{Q_1^{(n)}}}{q_E^2+m_{Q_1^{(n)}}^2}~.
\end{eqnarray}    
The quantities $F^{t_L,t_R}$ are the decay constants of the corresponding resonances. After expanding all the logarithms the potential can be written as follows,
\begin{equation}
\label{nmchm_B.3:7}
V_{\rm{eff}}(h,\eta)=-\frac{\mu_1^2}{2}h^2+\frac{\lambda_1}{4}h^4-\frac{\mu_2^2}{2}\eta^2+\frac{\lambda_2}{4}\eta^4-\frac{\lambda_m}{2}h^2\eta^2 ~,
\end{equation}
where the parameters $\mu_1,\lambda_1,\mu_2,\lambda_2$ and $\lambda_m$ are listed below:
\begin{eqnarray}
\label{nmchm_B.3:8}
\mu_1^2&=& 2\alpha_L-4s_\theta^2\alpha_R+4s_\theta^4\beta_R+2s_\theta^2\epsilon ~,\\
\label{nmchm_B.3:9}
\lambda_1&=&\beta_L+4s_\theta^4\beta_R+4s_\theta^2\epsilon ~,\\
\label{nmchm_B.3:10}
\mu_2^2&=& 4c_{2\theta}\alpha_R-4s_\theta^2c_{2\theta}\beta_R ~,\\
\label{nmchm_B.3:11}
\lambda_2&=& 4c_{2\theta}^2\beta_R ~,\\
\label{nmchm_B.3:12}
\lambda_m&=& 4s_\theta^2c_{2\theta}\beta_R+2c_{2\theta}\epsilon ~.
\end{eqnarray}
In the above expressions, $\alpha_{L,R},~\beta_{L,R}$ and $\epsilon$ denote the integrals over the momentum dependent form factors as follows: 
\begin{eqnarray}
\label{nmchm_B.3:13}
\alpha_{L,R}&=&N_c\int \frac{d^4q_E}{(2\pi)^4}\frac{\tilde{\Pi}^{t_L,t_R}_1}{\Pi^{t_L,t_R}_0} ~,
\\
\label{nmchm_B.3:14}
\beta_{L,R}&=& N_c\int \frac{d^4q_E}{(2\pi)^4} \left(\frac{\tilde{\Pi}^{t_L,t_R}_1}{\Pi^{t_L,t_R}_0}\right)^2 ,\\
\label{nmchm_B.3:15}
\epsilon&=& N_c\int \frac{d^4q_E}{(2\pi)^4}\frac{|M^{t}|^2}{q_E^2\Pi^{t_L}_0\Pi^{t_R}_0} ~.
\end{eqnarray}

It can be readily observed that, $\beta_{L,R}$ and $\epsilon$ become finite if $\tilde{\Pi}^{t_L,t_R}_1\sim\mathcal{O}(1/q_E^4)$, which can be achieved with two resonances only, whereas $\alpha_{L,R}$ become finite if $\tilde{\Pi}^{t_L,t_R}_1$  fall with $q_E$ faster than $\mathcal{O}(1/q_E^4)$. To achieve the latter, minimum three resonances are required. Note that, $\alpha_{L,R}$ are present only in the expressions for $\mu_1$ and $\mu_2$. Since the scalar mass matrix involves neither of the above mentioned two coefficients, employing only two resonances we can calculate the two scalar masses.  The masses of these two resonances (one singlet and another a five-plet of $\rm{SO(5)}$) have been denoted by $m_{Q_1}$ and $m_{Q_5}$. Assuming $\Pi^{t_L,t_R}_0\simeq 1$, the Weinberg sum rules arising from the condition, $\lim\limits_{q_E^2\rightarrow \infty} q_E^n\tilde{\Pi}^{t_L,t_R}_1=0$, with ($n=0,2$) lead to
\begin{equation}
\label{nmchm_B.3:16}
\left|F^{t_L,t_R}_1\right|=\left|F^{t_L,t_R}_5\right|\equiv\left|F^{t_L,t_R}\right|~.
\end{equation}
Imposing the above condition, the form factors turn out to be
\begin{eqnarray}
\label{nmchm_B.3:17}
\tilde{\Pi}^{t_L,t_R}_1=\frac{\left|F^{t_L,t_R}\right|^2(m_{Q_5}^2-m_{Q_1}^2)}{(q_E^2+m_{Q_1}^2)(q_E^2+m_{Q_5}^2)} ~.
\end{eqnarray}  
To calculate $M^t$, we make another assumption that $F^{t_L,t_R}_1$ are real, and 
\begin{equation}
\label{phase}
F^{t_L}_5F^{t_R*}_5\simeq \left|F^{t_L}\right|\left|F^{t_R}\right|e^{i\theta_{\rm{phase}}}~.
\end{equation}
With the above assumption 
\begin{eqnarray}
\nonumber
M^t=\frac{\left|F^{t_L}\right|\left|F^{t_R}\right|}{(q_E^2+m_{Q_1}^2)(q_E^2+m_{Q_5}^2)}\left[(m_{Q_1}-m_{Q_5}e^{i\theta_{\rm{phase}}})q_E^2\right.\\
\label{nmchm_B.3:18}
\left. +m_{Q_1}m_{Q_5}(m_{Q_5}-m_{Q_1}e^{i\theta_{\rm{phase}}})\right]~.
\end{eqnarray}
Employing all these relations, the integrals $\beta_{L,R}$ and $\epsilon$ can be evaluated exactly as
\begin{eqnarray}
\label{nmchm_B.3:19}
\beta_{L,R}=\frac{N_c}{8\pi^2}\left|F^{t_L,t_R}\right|^4\left[\frac{m_{Q_1}^2+m_{Q_5}^2}{2(m_{Q_1}^2-m_{Q_5}^2)}\log\left(\frac{m_{Q_1}^2}{m_{Q_5}^2}\right)-1\right]~,
\end{eqnarray}
and
\begin{eqnarray}
\label{nmchm_B.3:20}
\epsilon=\frac{N_c}{8\pi^2}\left|F^{t_L}\right|^2\left|F^{t_R}\right|^2\left[1-\cos\theta_{\rm{phase}}\frac{m_{Q_1}m_{Q_5}}{m_{Q_1}^2-m_{Q_5}^2}\log\left(\frac{m_{Q_1}^2}{m_{Q_5}^2}\right)\right]~.
\end{eqnarray}

\bibliographystyle{JHEP}
\bibliography{FTcomp.bib}

\end{document}